# A new efficient Hamiltonian approach to the nonlinear water-wave problem over arbitrary bathymetry

Ch.E. Papoutsellis ([1]), G.A. Athanassoulis ([2]) ([3]),

**Abstract**

A new Hamiltonian formulation for the fully nonlinear water-wave problem over variable bathymetry is derived, using an exact, vertical series expansion of the velocity potential, in conjunction with Luke's variational principle. The obtained Euler-Lagrange equations contain infinite series and can rederive various existing model equations upon truncation. In this paper, the infinite series are summed up, resulting in two exact Hamiltonian equations for the free-surface elevation and the free-surface potential, coupled with a time-independent horizontal system of equations. The Dirichlet to Neumann operator is given by a simple and versatile representation, which is valid for any smooth fluid domain, not necessarily periodic in the horizontal direction(s), without limitations on the steepness and deformation of the seabed and the free surface. An efficient numerical scheme is presented and applied to the case of one horizontal dimension, establishing the ability of the new formulation to simulate strongly nonlinear and dispersive wave-bottom interactions by comparison with experimental measurements.

**Table of contents**

([1]) PhD, National Technical University of Athens, School of Naval Architecture and Marine Engineering, Athens, Greece, Email: cpapoutse@central.ntua.gr, cpapoutsellis@gmail.com

([2]) Professor, National Technical University of Athens, School of Naval Architecture and Marine Engineering, Athens, Greece, Email: mathan@central.ntua.gr, makathan@gmail.com

([3]) Research Professor, Research Center for High Performance Computing, ITMO University, St. Petersburg, Russia

## 1. Introduction

The mathematical modelling of water-waves is a demanding task even under the simplifying physical context of an inviscid fluid under irrotational flow. In fact, the presence of the unknown free surface makes the mathematical formulation of the hydrodynamic problem nonlinear and nonlocal. Additional complexity is introduced, when water-waves are propagating over a non-flat seabed, in which case, the accurate description of the wave motion is of major importance for the study of the complex coastal environment and its implications on modern engineering applications.

The classical mathematical formulation of the nonlinear, irrotational, water-wave problem (NLIWW problem) possesses a nonlocal Hamiltonian structure, discovered by (Zakharov, 1968). The Hamiltonian of the problem is the total energy of the fluid, considered as a functional on the free-surface elevation $\eta = \eta(\boldsymbol{x}, t)$ and the trace of the wave potential on the free surface, $\psi(\boldsymbol{x}, t) = \Phi(\boldsymbol{x}, \eta(\boldsymbol{x}, t), t)$, under the constraint that the potential $\Phi(\boldsymbol{x}, z, t)$ satisfies the interior fluid kinematics, at every time instant $t$; see also (Miles, 1977), (Milder, 1977), (Benjamin & Olver, 1982). As pointed out by Craig & Sulem (1993), the closure of Zakharov's equations requires the knowledge of the Dirichlet to Neumann (DtN) operator associated with the solution of the boundary-value problem (BVP) corresponding to internal kinematics.

Exploiting the analyticity of either the wave potential or the DtN operator, many authors developed pertubative procedures for the calculation of surface quantities required for the closure of the dynamical equations (free-surface conditions). In the flat-bottom case, and assuming horizontal periodicity, (Dommermuth & Yue, 1987) and (West & Brueckner, 1987) approximated the potential using perturbation series, while Craig & Sulem (1993) proposed the approximation of the DtN operator by an operator expansion, in the form of a functional Taylor (Volterra-Taylor) series. Further variants and developments of the latter approach can be found in Nicholls (1998), Bateman *et al.* (2001), Craig & Nicholls (2002), (Nicholls & Reitich, 2001a), (Nicholls & Reitich, 2001b), (Nicholls, 2007), see also (Schäffer, 2008). Expansions of such type have been generalized to variable bathymetry, assuming that the seabed surface is a small perturbation of the horizontal plane, by (Smith, 1998), (Guyenne & Nicholls, 2005), (Craig, Guyenne, Nicholls, & Sulem, 2005), (Craig, Guyenne, & Sulem, 2009), (Gouin, Ducrozet, & Ferrant, 2016). Pertubative approaches lead to numerical schemes of low computational cost, being efficient and reliable for reasonable steepness of the bathymetry and the surface waves. For stable long time-simulations, filtering of intermediate results is often required (Dommermuth & Yue, 1987), (Xu & Guyenne, 2009). It is interesting to note that (Milder, 1990) and (Ambrose, Bona, Nicholls, 2014), using simplified models, conclude that the truncation of the Hamiltonian seems to result in artificial instabilities.

An approach that avoids the use (and the limitations) of perturbation series, is to evolve the free-surface boundary conditions, while solving the instantaneous kinematic BVP for the wave potential by using direct numerical methods, e.g., the finite-element method Gagarina *et al.* (2013), or the finite-difference method Bingham & Zhang (2007), or boundary integral equations (Grilli, 1996), Clamond & Grue (2001), (Grilli, Guyenne, & Dias, 2001). In general, such methods provide high-quality numerical results without limitations on the geometry of the fluid domain. However, their implementation in domains of large horizontal extend becomes slow, and sophisticated acceleration techniques are required. Another approach that can, in principle, treat the fully nonlinear problem has been developed by (Ablowitz, Fokas, & Musslimani, 2006); see also (Bridges, Groves, & Nicholls, 2016), Chapter 5. In the special case of one horizontal dimension, an effective alternative is possible, by exploiting complex analytic functions and conformal mapping techniques; see e.g. (Longuet-Higgins & Cokelet, 1975), (Zakharov, Dyachenko, & Vasilyev, 2002), (Chalikov & Sheinin, 2005), (Viotti, Dutykh, & Dias, 2013). Interesting comparative studies of some of the above methods can be found in (Wilkening & Vasan, 2015) and ( Zhang, Sanina, Babanin, & Guedes Soares, 2016).

The demand of fast long-time simulations with adequate physical accuracy, led also to the development of a plethora of model equations, understood as improvements of the original Boussinesq's equations. A great deal of progress has been made in deriving Boussinesq type models (BTMs) with improved dispersion characteristics. We mention, for example, the recent developments by (Bingham & Agnon, 2005) (Madsen, Fuhrman, & Wang, 2006), (Karambas & Memos, 2009) and refer to (Brocchini, 2013), (Memos, Klonaris, & Chondros, 2015) and (Ma, 2010), Chapter 7, for extended reviews and further references. Recently, attention has also been paid on another class of horizontally reduced local models, the Green-Naghdi (GN) equations, first appeared in 70s' (Green & Naghdi, 1976) and almost forgotten for more than 20 years; see e.g., (Kim, Bai, Ertekin, & Webster, 2001), (Didier Clamond & Dutykh, 2011), (Webster, Duan, & Zhao, 2011), (Bonneton et al., 2011), (Zhao, Duan, & Ertekin, 2014), (Matsuno, 2016), (D Clamond, Dutykh, & Mitsotakis, 2017). In general, the extended region of validity of BTMs and GN models is redeemed by the presence of high-order polynomial nonlinearities and high-order derivatives in the evolution equations.

Several authors also derived model equations based on approximations of the velocity potential by series representations in terms of vertical functions and undetermined horizontal coefficients. (Isobe & Abohadima, 1998) and Klopman *et al.* (2010) worked with ad hoc vertical series approximations and mild-slope assumptions. Tian and Sato (2008) approximated the vertical structure of the velocity potential using Chebyshev polynomials, obtaining a promising model. This approach has been further developed and validated by (Yates & Benoit, 2015), (Raoult, Benoit, & Yates, 2016).

Our main goal in this paper is to derive a new, efficient Hamiltonian formulation for the exact nonlinear water-wave problem over varying bathymetry, by avoiding both perturbative techniques and 3D direct numerical methods. We start from the exact representation of the wave potential $\Phi(\boldsymbol{x},z,t)$ in the instantaneous fluid domain, by using a rapidly convergent series expansion, first introduced by Athanassoulis & Belibassakis (2000), (Athanassoulis & Belibassakis, 2007), and studied in detail by (Athanassoulis & Papoutsellis, 2017) (referred to as AP17 in the following):

$$\Phi(\boldsymbol{x},z,t) = \sum_{n} \varphi_n(\boldsymbol{x},t)\, Z_n\big(z\,;\eta(\boldsymbol{x},t),\, h(\boldsymbol{x})\big),$$

where $h(\boldsymbol{x})$ is the parametrization of the seabed, $Z_n\big(z;\eta(\boldsymbol{x},t),h(\boldsymbol{x})\big)$ are vertical functions consisting a basis in the Sobolev space $H^2\big([-h(\boldsymbol{x}),\eta(\boldsymbol{x},t)]\big)$, and $\varphi_n(\boldsymbol{x},t)$ are unknown coefficients (*modal amplitudes*). As shown in AP17, this expansion converges rapidly and can be term-wise differentiated two times, up to and including the boundaries, provided that the latter are smooth enough. Using this expansion, in conjunction with Luke's variational principle (Luke, 1967), we obtain, as Euler-Lagrange (EL) equations, horizontal differential equations for $\eta(\boldsymbol{x},t)$ and $\varphi_n(\boldsymbol{x},t)$, which constitute an exact reformulation of the full problem. An unpleasant feature of these EL equations is that they contain infinite series, summed over all unknown modal amplitudes $\varphi_n(\boldsymbol{x},t)$. It is possible to truncate the infinite series and introduce mild-slope assumptions, obtaining various model equations, e.g. the ones presented by Isobe and Abohadima (1998) and Klopman *et al.* (2010), as well as all versions of the consistent coupled-mode systems (CCMS), as e.g. Athanassoulis & Belibassakis (1999), Athanassoulis & Belibassakis (2007), Belibassakis & Athanassoulis (2011). This possibility, establishing consistency with previous approaches, is briefly discussed before proceeding to our principal goal, which is to derive a more convenient, exact reformulation. This is accomplished by exploiting the convergence properties of the exact series expansion, which permit us to sum up the infinite series, obtaining two simple Hamiltonian equations for $\eta=\eta(\boldsymbol{x},t)$ and $\psi=\psi(\boldsymbol{x},t)$, coupled with a system of time-independent, horizontal, partial differential equations (substrate system), that determines the modal amplitudes $\varphi_n(\boldsymbol{x},t)$ at every time $t$. This new formulation is free of any restrictions concerning nonlinearity, shallowness, bottom variation and horizontal periodicity. The two nonlinear evolution equations contain only first-order derivatives in space and time, while the nonlocality of the problem is exactly represented by means of a single modal-amplitude field. In this way, the DtN operator of the fully nonlinear problem, with varying bathymetry, is accurately and efficiently modelled and calculated at every time $t$ via a genuine dimensionally reduced procedure.

The organization of the paper is as follows. In Section 2, the classical differential formulation and the corresponding Luke's variational principle of the fully nonlinear problem are briefly presented. In Section 3, we derive the EL equations corresponding to a generic series representation of the velocity potential and briefly discuss the (re)derivation of some existing models from these equations. In Section 4, we exploit the exactness and the specific structure of the vertical series expansion given in AP17 to derive our new Hamiltonian formulation. In Section 5, a numerical scheme for the solution of the unidirectional version of our equations is introduced, applied and validated against experiments involving highly non-linear and dispersive waves over mild and steep varying bathymetry. Finally, in Section 6 the main findings are summarized and discussed.

## 2. Formulation of the problem

Let $O\boldsymbol{x}z$, $\boldsymbol{x}=(x_1,x_2)$, be an orthogonal Cartesian co-ordinate system, with vertical $z-$axis pointing upward. The still water level corresponds to $z=0$. The unknown, time-dependent, fluid domain $D_h^\eta(X,t)\subset\mathbb{R}^{d+1}$, $d=1$ or $2$, is defined by

$$D_h^\eta(X,t)=\left\{(\boldsymbol{x},z)\in X\times\mathbb{R}:\boldsymbol{x}\in X,-h(\boldsymbol{x})<z<\eta(\boldsymbol{x},t)\right\}. \quad (2.1)$$

where $X$ denotes the horizontal fluid extent, and the (single-valued) smooth functions $h:X\to\mathbb{R}$ and $\eta:X\times[t_0,t_1]\to\mathbb{R}$ denote the (finite) water depth and the free-surface elevation, respectively. We also introduce the *excitation boundary*, defined by $S_a=\left\{(\boldsymbol{x},z):x_1=a,\,x_2\in\mathbb{R},\,-h_a\le z\le\eta(a,x_2,t)\right\}$, where $h_a$ and $\eta(a,x_2,t)$ are, respec-tively, the depth([4]) and the free-surface elevation along the boundary line $\{x_1 = a,\, x_2 \in \mathbb{R}\}$ of the half plane $X$ ; see **Figure 1**. The mathematical formulation of the nonlinear irrotational water waves (NLIWW) problem, is stated in terms of the free-surface elevation $\eta = \eta(\boldsymbol{x}, t)$ and the velocity potential $\Phi = \Phi(\boldsymbol{x}, z, t)$ and includes the following equations (see e.g. Stoker (1957), Wehausen (1960)):

$$\Delta\Phi \;=\; \nabla_{\boldsymbol{x}}^2\Phi + \partial_z^2\Phi \;=\; 0\,, \quad \text{on} \quad D_h^\eta(X,t)\,, \;\; t_0 \le t \le t_1\,, \tag{2.2a}$$

$$\boldsymbol{N}_h \cdot \nabla\Phi \;=0\,, \qquad \text{on} \quad \Gamma_h(X)\,, \;\; t_0 \le t \le t_1\,, \tag{2.2b}$$

$$\partial_t\eta \,-\, \boldsymbol{N}_\eta \cdot \nabla\Phi = \,0\,, \qquad \text{on} \quad \Gamma^\eta(X,t)\,, \;\; t_0 \le t \le t_1\,, \tag{2.2c}$$

$$\partial_t\Phi + \frac{1}{2}(\nabla\Phi)^2 + g\eta + \frac{p_{\text{surf}}}{\rho} = 0\,, \quad \text{on} \quad \Gamma^\eta(X,t)\,, \;\; t_0 \le t \le t_1\,, \tag{2.2d}$$

where $\nabla_{\boldsymbol{x}} = (\partial_{x_1}, \partial_{x_2})$ is the horizontal (2D) gradient, $\nabla = (\nabla_{\boldsymbol{x}}, \partial_z) = (\partial_{x_1}, \partial_{x_2}, \partial_z)$ is the usual three-dimensional (3D) gradient, $\nabla_{\boldsymbol{x}}^2 = (\partial_{x_1}^2, \partial_{x_2}^2)$ is the horizontal Laplacian, $\partial_t$ denotes the time derivative, $\boldsymbol{N}_h = (-\nabla_{\boldsymbol{x}} h, -1)$ and $\boldsymbol{N}_\eta = (-\nabla_{\boldsymbol{x}}\eta, 1)$ are the outward (with respect to the fluid) normal (but *not normalized* to have unit length) vectors on $\Gamma_h(X)$ and $\Gamma^\eta(X,t)$, respectively, and $p_{\text{surf}}(\boldsymbol{x},t)$ denotes any externally applied pressure. Further, $g$ is the acceleration due to gravity, and $\rho$ is the (constant) density of the fluid. The local total depth $H(\boldsymbol{x},t) = \eta(\boldsymbol{x},t) + h(\boldsymbol{x})$ is assumed smooth and bounded:

$$0 < H_{\min} \le H(\boldsymbol{x},t) \le H_{\max} \quad \text{in} \quad X \times [t_0\,, t_1]\,.$$

We also assume that $\eta(\boldsymbol{x},t)$ and $\nabla\Phi(\boldsymbol{x},z,t)$ vanish at infinity, as $|\boldsymbol{x}| \to \infty$. Equations (2.2a-d) should be supplemented by appropriate conditions on the lateral boundaries:

$$\eta(a, x_2, t) = \eta_a(x_2, t) \equiv \eta_a\,, \qquad x_2 \in \mathbb{R}\,, \quad t_0 \le t \le t_1\,, \tag{2.3a}$$

$$\left\{\begin{array}{ll} \text{either} & \partial_{x_1}\Phi(a,x_2,z,t) = V_a(x_2,z,t) \equiv V_a \\ \text{or} & \Phi(a,x_2,z,t) = \Phi_a(x_2,z,t) \equiv \Phi_a \end{array}\right\}, \quad \boldsymbol{x} \in S_a\,, \quad t_0 \le t \le t_1\,, \qquad \begin{array}{r}(2.3b)\\(2.3c)\end{array}$$

Equations (2.3a-c) specify, respectively, the free-surface elevation and the values of the normal velocity (or, alternatively, the values of the wave potential) on the excitation boundary $S_a$ at each time instance $t$. The data $\eta_a$ and $V_a$ (or $\Phi_a$) are known functions, obtained by using, e.g., a (precalculated) progressive nonlinear wave at the constant depth $h_a$, or a simulated stochastic wave field.

([4]) The depth is assumed constant in the vicinity of the excitation boundary. The theory developed herein can be trivially extended to any shape of vertical lateral boundaries.

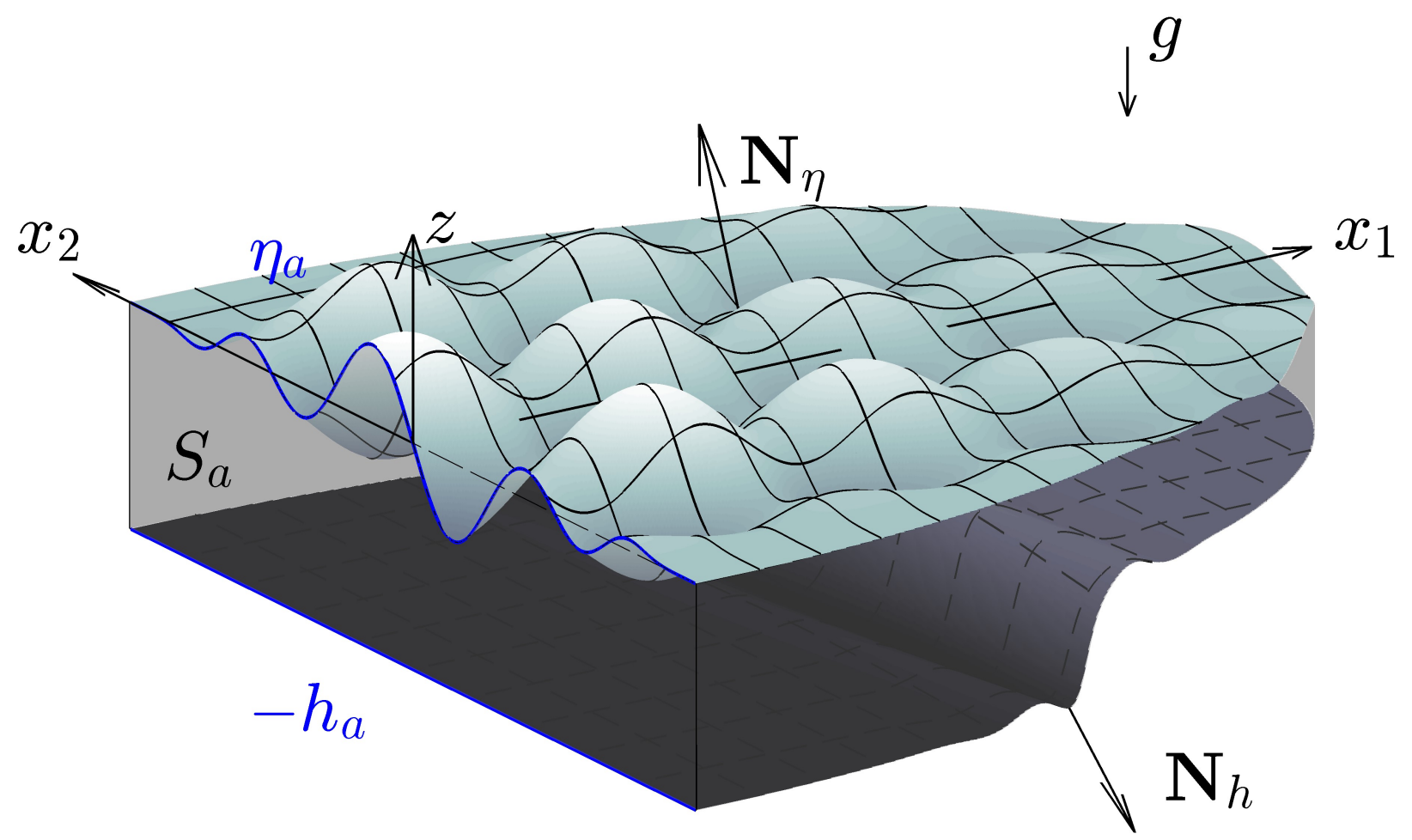

**Figure 1.** The geometric configuration of the fluid domain

## 2.1 *Luke's Variational formulation*

Luke (1967) provided an unconditional variational principle for the NLIWW problem, Eqs. (2.2), for the case of an horizontally unbounded fluid and $p_{\text{surf}} = 0$. Luke's Lagrangian density, augmented by two simple terms accounting for the presence of applied pressure and lateral excitation, is given below:

$$\mathcal{L}(\eta,\Phi) = \mathcal{L}(\eta(\boldsymbol{x},t),\Phi(\boldsymbol{x},\cdot,t)) =$$

$$= \int_{-h}^{\eta} \left( \partial_t \Phi + \frac{1}{2}|\nabla_{\boldsymbol{x}} \Phi|^2 + \frac{1}{2}(\partial_z \Phi)^2 + g\,z \right) dz + \frac{p_{\text{surf}}}{\rho}\eta + \begin{cases} \displaystyle\int_{-h_a}^{\eta_a} V_a \left[ \Phi \right]_{x_1 = a} dz \\ 0 \end{cases} \tag{2.4}$$

The two different possibilities appearing in the last term of Eq. (2.4) reflect the fact that excitation of the fluid domain may occur either by specifying the normal velocity $V_a$ or the values the potential $\Phi_a$ there. In the sequel, we shall explicitly treat the first case. The corresponding action functional is defined by integrating $\mathcal{L}[\eta,\Phi]$ over the region $X \times [t_0, t_1]$:

$$\mathcal{S}[\eta,\Phi] = \int_{t_0}^{t_1} \int_X \mathcal{L}(\eta,\Phi)\, d\boldsymbol{x}\, dt\,. \tag{2.5}$$

Luke's variational principle states that the two fields $\eta, \Phi$ satisfy the NLIWW problem, Eqs (2.2), (2.3), if and only if, they render the functional $\mathcal{S}[\eta,\Phi]$ stationary, that is,

$$\delta\mathcal{S}[\eta,\Phi;\delta\eta,\delta\Phi] = \delta_\Phi \mathcal{S}[\eta,\Phi;\delta\Phi] + \delta_\eta \mathcal{S}[\eta,\Phi;\delta\eta] = 0, \tag{2.6}$$

for all admissible variations $\delta\eta, \delta\Phi$. Here $\delta_\Phi \mathcal{S}[\eta,\Phi;\delta\Phi]$ and $\delta_\eta \mathcal{S}[\eta,\Phi;\delta\eta]$ denote the first partial variations (functional derivatives) of $\mathcal{S}[\eta,\Phi]$ with respect to $\Phi$ and $\eta$ in the directions $\delta\Phi$ and $\delta\eta$, respectively. Their calculation, for the case of a horizontally unbounded fluid, has been given by Luke (1967) (see also (Whitham, 1974; Sec. 13.2.)). They are given here, including also the effects of the lateral boundary condition and the applied pressure, for further use in Section 3:

$$\delta_\Phi \mathcal{S}[\eta,\Phi;\delta\Phi] = \int_{t_0}^{t_1}\int_X \Big\{ \big(-\partial_t\eta + \boldsymbol{N}_\eta\cdot[\nabla\Phi]_{z=\eta}\big)[\delta\Phi]_{z=\eta} - \int_{-h}^{\eta}\Delta\Phi\,\delta\Phi\,dz + \boldsymbol{N}_h\cdot[\nabla\Phi\,\delta\Phi]_{z=-h}\Big\}\,d\boldsymbol{x}\,dt + \int_{t_0}^{t_1}\int_{\mathbb{R}}\int_{-h_a}^{\eta_a}\Big(-\big[\partial_{x_1}\Phi\big]_{x_1=a} + V_a\Big)[\delta\Phi]_{x_1=a}\,dz\,dx_2\,dt, \tag{2.7a}$$

$$\delta_\eta \mathcal{S}[\eta,\Phi;\delta\eta] = \int_{t_0}^{t_1}\int_X \Big\{[\partial_t\Phi]_{z=\eta} + \frac{1}{2}[\nabla\Phi]^2_{z=\eta} + g\eta + \frac{p_{\text{surf}}}{\rho}\Big\}\,\delta\eta\,d\boldsymbol{x}\,dt. \tag{2.7b}$$

In Eqs (2.7), as well as in the sequel, boundary values (traces) of the various fields will be denoted by using brackets with a subscript specifying the boundary.

## 3. Variational reformulation using a generic vertical series representations for the wave potential

The usefulness of Luke's variational principle, Eq. (2.7), lies on the fact that, being unconstrained, it allows one to use any convenient representation of the wave potential $\Phi = \Phi(\boldsymbol{x},z,t)$, avoiding the a priori consideration of kinematics, described by Eqs. (2.2a,b). This possibility has already been exploited by many authors for the development of simplified water-wave models, based on ad hoc series approximation of the wave potential, of the form

$$\Phi(\boldsymbol{x},z,t) = \sum_n \varphi_n(\boldsymbol{x},t)\,Z_n(z;\eta,h), \tag{3.1}$$

where $Z_n$ are explicitly prescribed functions of $z$ (vertical functions), possibly dependent on the local values of $\eta(\boldsymbol{x};t)$ and $h(\boldsymbol{x})$, and $\varphi_n(\boldsymbol{x},t)$ are unknown (to be determined) modal amplitudes. Usually, $Z_n$ are chosen as eigenfunctions of the linear problem, Massel (1993), Porter & Staziker (1995), (Kaihatu & Kirby, 1995), (Chamberlain & Porter, 2006), or polynomials (sometimes in combination with hyperbolic functions) (Kim et al., 2001), (Klopman et al., 2010), (Isobe & Abohadima, 1998), (Isobe, 2013; Sec. 4.5). Another variant of Eq. (3.1), initially proposed by (Athanassoulis & Belibassakis (1999), Athanassoulis & Belibassakis (2000)), and extensively used later on, (Belibassakis, Athanassoulis, & Gerostathis, 2001), (Belibassakis & Athanassoulis, 2002), (Athanassoulis & Belibassakis, 2007), (Belibassakis & Athanassoulis, 2011)), combines the eigenfunctions expansion with additional modes, accounting for the boundary conditions at $z = -h(\boldsymbol{x})$ and $z = \eta(\boldsymbol{x},t)$. The obtained EL equations will be either approximate model equations or exact reformulation of the problem, depending on the character (approximate or exact) of Eq. (3.1). In this section, we present a systematic derivation of the EL equations for the fields $\eta$ and $\varphi_n$ by applying Luke's variational principle in conjunction with Eq. (3.1), either approximate or exact. The advantages of using the exact series expansion developed in AP17 are discussed in the next section.

Introducing the notation $\Phi = \sum \varphi_n Z_n = \boldsymbol{\varphi}^{\mathrm{T}} \boldsymbol{Z}$, where $\boldsymbol{Z} = \boldsymbol{Z}(z,\eta,h) \equiv \{Z_n(z;\eta,h)\}_n$ is the sequence of vertical functions and $\boldsymbol{\varphi} \equiv \boldsymbol{\varphi}(\boldsymbol{x},t) = \{\varphi_n(\boldsymbol{x},t)\}_n$ is the sequence of modal amplitudes, the action functional (2.5) is reformulated as a functional on $(\eta,\boldsymbol{\varphi})$, taking the form

$$\tilde{\mathcal{S}}[\eta,\boldsymbol{\varphi}] = \int_{t_0}^{t_1} \left\{ \int_X \left[ \int_{-h}^{\eta} \left( \partial_t(\boldsymbol{\varphi}^{\mathrm{T}} \boldsymbol{Z}(\eta)) + \frac{1}{2} |\nabla(\boldsymbol{\varphi}^{\mathrm{T}} \boldsymbol{Z}(\eta))|^2 \right) dz + \frac{1}{2} g \eta^2 + \frac{p_{\mathrm{surf}}}{\rho} \eta \right] d\boldsymbol{x} \right\} dt$$

$$+ \int_{t_0}^{t_1} \int_{\mathbb{R}} \int_{-h_a}^{\eta_a} V_a \left[ \boldsymbol{\varphi}^{\mathrm{T}} \boldsymbol{Z}(\eta) \right]_{x_1 = a} dz\, dx_2\, dt. \quad (3.2)$$

The EL equations in the new functional variables $(\eta,\boldsymbol{\varphi})$ are obtained from the variational equation

$$\delta\tilde{\mathcal{S}}[\eta,\boldsymbol{\varphi};\delta\eta,\delta\boldsymbol{\varphi}] = \delta_\eta \tilde{\mathcal{S}}[\eta,\boldsymbol{\varphi};\delta\eta] + \sum_m \delta_{\varphi_m} \tilde{\mathcal{S}}[\eta,\boldsymbol{\varphi};\delta\varphi_m] = 0, \quad (3.3)$$

where $\delta_\eta \tilde{\mathcal{S}}[\eta,\boldsymbol{\varphi};\delta\eta]$ and $\delta_{\boldsymbol{\varphi}} \tilde{\mathcal{S}}[\eta,\boldsymbol{\varphi};\delta\boldsymbol{\varphi}]$ are the partial variations (functional derivatives) of $\tilde{\mathcal{S}}[\eta,\boldsymbol{\varphi}]$ in the directions $\delta\eta$ and $\delta\boldsymbol{\varphi}$. The variations $(\delta\eta,\delta\boldsymbol{\varphi})$ are arbitrary, admissible (continuously differentiable) functions of $(\boldsymbol{x},t)$ that satisfy the isochronality constraint (they vanish at $t = t_0, t_1$), the rest-at-infinity condition ($\delta\eta,\delta\varphi_n \to 0$ as $|\boldsymbol{x}| \to \infty$), and the condition $[\delta\eta]_{x_1 = a} = 0$, due to the fact that the free-surface elevation is assumed known on the excitation boundary. To calculate the EL equations for $(\eta,\boldsymbol{\varphi})$ we need to calculate the variation of $\tilde{\mathcal{S}}$ as Eq. (3.3) dictates. Although this calculation can be done directly by expanding the right-hand side of Eq. (3.2) and performing straightforward (yet laborious) manipulations, it is more instructive and more economical to exploit the fact that $\tilde{\mathcal{S}}$ is a composite functional and apply the chain rule for functional derivatives (see, e.g., (Flett, 1980), 4.1.2, (Gasinski & Papageorgiou, 2005), Prop. 4.1.12). In this way, we are able to utilize the already calculated variations $\delta_\Phi \mathcal{S}[\eta,\Phi;\delta\Phi]$ and $\delta_\eta \mathcal{S}[\eta,\Phi;\delta\eta]$, given by Eqs. (2.7a) and (2.7b), respectively, and keep a better track of the various terms arising from the variational procedure. This is accomplished by the following lemma, which is proved in Appendix A.

**Lemma 1:** The following formulae hold true

$$\delta_\eta \tilde{\mathcal{S}}[\eta,\boldsymbol{\varphi};\delta\eta] = \delta_\eta \mathcal{S}[\eta,\boldsymbol{\varphi}^{\mathrm{T}}\boldsymbol{Z}(\eta);\delta\eta] + \delta_\Phi \mathcal{S}[\eta,\boldsymbol{\varphi}^{\mathrm{T}}\boldsymbol{Z}(\eta);(\boldsymbol{\varphi}^{\mathrm{T}}\partial_\eta \boldsymbol{Z}(\eta))\delta\eta], \quad (3.4a)$$

$$\delta_{\varphi_m} \tilde{\mathcal{S}}[\eta,\boldsymbol{\varphi};\delta\varphi_m] = \delta_\Phi \mathcal{S}[\eta,\boldsymbol{\varphi}^{\mathrm{T}}\boldsymbol{Z}(\eta);Z_m(\eta)\delta\varphi_m]. \quad (3.4b)$$

Let us start by calculating the variation $\delta_{\varphi_m} \tilde{\mathcal{S}}[\eta,\boldsymbol{\varphi};\delta\varphi_m]$. Invoking (3.4b), we substitute $\Phi$ and $\delta\Phi$ in Eq. (2.7a), by $\boldsymbol{\varphi}^{\mathrm{T}}\boldsymbol{Z}(\eta)$ and $Z_m(\eta)\delta\varphi_m$. Factoring out $\delta\varphi_m$ and $[\delta\varphi_m]_{x_1=a}$ from the vertical integral, we obtain

$$\delta_{\varphi_m}\tilde{\mathcal{S}}[\eta,\boldsymbol{\varphi};\delta\varphi_m]=\int_{t_0}^{t_1}\Bigg\{\int_X\Bigg[\Big(-\partial_t\eta+\boldsymbol{N}_\eta\cdot\big[\nabla(\boldsymbol{\varphi}^{\mathrm{T}}\boldsymbol{Z})\big]_{z=\eta}\Big)\big[Z_m\big]_{z=\eta}$$

$$-\int_{-h}^{\eta}\Delta(\boldsymbol{\varphi}^{\mathrm{T}}\boldsymbol{Z})Z_m\,dz-\boldsymbol{N}_h\cdot\big[\nabla(\boldsymbol{\varphi}^{\mathrm{T}}\boldsymbol{Z})Z_m\big]_{z=-h}\Bigg]\delta\varphi_m\,d\boldsymbol{x} \tag{3.5a}$$

$$+\int_{\mathbb{R}}\Bigg[\int_{-h_a}^{\eta_a}\Big(\big[\partial_{x_1}(\boldsymbol{\varphi}^{\mathrm{T}}\boldsymbol{Z})\big]_{x_1=a}-V_a\Big)[Z_m]_{x_1=a}\,dz\Bigg]\big[\delta\varphi_m\big]_{x_1=a}dx_2\Bigg\}dt,$$

where, for the sake of brevity, the $\eta$ dependence of $\boldsymbol{Z}(\eta)$ has been dropped. Recalling the notation $\nabla=(\nabla_{\boldsymbol{x}},\partial_z)$, $\Delta=\nabla_{\boldsymbol{x}}^2+\partial_z^2$, $\boldsymbol{N}_h=(-\nabla_{\boldsymbol{x}}h,-1)$, and performing term-wise differentiations and integrations on the series $\boldsymbol{\varphi}^{\mathrm{T}}\boldsymbol{Z}$, we obtain the equations

$$-\int_{-h}^{\eta}\Delta(\boldsymbol{\varphi}^{\mathrm{T}}\boldsymbol{Z})Z_m\,dz-\boldsymbol{N}_h\cdot\big[\nabla(\boldsymbol{\varphi}^{\mathrm{T}}\boldsymbol{Z})Z_m\big]_{z=-h}=$$

$$=-\sum_n\Bigg\{\Big(\int_{-h}^{\eta}Z_nZ_m\,dz\Big)\nabla_{\boldsymbol{x}}^2\varphi_n+\Big(2\int_{-h}^{\eta}(\nabla_{\boldsymbol{x}}Z_n)Z_m\,dz+(\nabla_{\boldsymbol{x}}h)\big[Z_mZ_n\big]_{z=-h}\Big)\cdot\nabla_{\boldsymbol{x}}\varphi_n$$

$$+\Big(\int_{-h}^{\eta}(\nabla_{\boldsymbol{x}}^2Z_n+\partial_z^2Z_n)Z_m\,dz-\boldsymbol{N}_h\cdot\big[(\nabla_{\boldsymbol{x}}Z_n,\partial_zZ_n)Z_m\big]_{z=-h}\Big)\varphi_n\Bigg\}, \tag{3.5b}$$

$$\int_{-h_a}^{\eta_a}\Big(\big[\partial_{x_1}(\boldsymbol{\varphi}^{\mathrm{T}}\boldsymbol{Z})\big]_{x_1=a}-V_a\Big)[Z_m]_{x_1=a}\,dz=\Big(\int_{-h_a}^{\eta_a}[Z_nZ_m]_{x_1=a}\,dz\Big)\big[\partial_{x_1}\varphi_n\big]_{x_1=a}+$$

$$+\Big(\int_{-h_a}^{\eta_a}\big[(\partial_{x_1}Z_n)Z_m\big]_{x_1=a}\,dz\Big)[\varphi_n]_{x_1=a}-\int_{-h_a}^{\eta_a}V_a[Z_m]_{x_1=a}\,dz. \tag{3.5c}$$

To make the form of the obtained expressions concise, we introduce the differential operators

$$L_{mn}[\eta,h]\bullet=A_{mn}(\eta,h)\nabla_{\boldsymbol{x}}^2\bullet+\boldsymbol{B}_{mn}(\eta,h)\cdot\nabla_{\boldsymbol{x}}\bullet+C_{mn}(\eta,h)\bullet, \tag{3.6a}$$

$$T_{mn}\bullet=\big[A_{mn}\big]_{x_1=a}\big[\partial_{x_1}\bullet\big]_{x_1=a}+\frac{1}{2}\big[B_{mn}^{(1)}\big]_{x_1=a}[\bullet]_{x_1=a}, \tag{3.6b}$$

defined respectively on $X$ and $\{x_1=a\}$, with

$$A_{mn}=\int_{-h}^{\eta}Z_nZ_m\,dz, \tag{3.7a}$$

$$\boldsymbol{B}_{mn}=(B_{mn}^{(1)},B_{mn}^{(2)})=2\int_{-h}^{\eta}(\nabla_{\boldsymbol{x}}Z_n)Z_m\,dz+(\nabla_{\boldsymbol{x}}h)\big[Z_mZ_n\big]_{z=-h}, \tag{3.7b}$$

$$C_{mn}=\int_{-h}^{\eta}(\nabla_{\boldsymbol{x}}^2Z_n+\partial_z^2Z_n)Z_m\,dz-\boldsymbol{N}_h\cdot\big[(\nabla_{\boldsymbol{x}}Z_n,\partial_zZ_n)Z_m\big]_{z=-h}, \tag{3.7c}$$

and note the useful identities

$$\sum_n L_{mn}[\eta,h]\varphi_n=\int_{-h}^{\eta}\Delta(\boldsymbol{\varphi}^{\mathrm{T}}\boldsymbol{Z})Z_m\,dz-\boldsymbol{N}_h\cdot\big[\nabla(\boldsymbol{\varphi}^{\mathrm{T}}\boldsymbol{Z})Z_m\big]_{z=-h}, \tag{3.8a}$$

$$\sum_n T_{mn}\,\varphi_n = \int_{-h_a}^{\eta_a} \left[\partial_{x_1}(\boldsymbol{\varphi}^{\mathrm{T}}\boldsymbol{Z})\right]_{x_1=a} \left[Z_m\right]_{x_1=a} dz\,. \tag{3.8b}$$

(Note that we have made use of the assumption $\left[\partial_{x_1} h\right]_{x_1=a} = 0$, for the derivation of Eq. (3.8b)). Term-wise differentiation is unconditionally permitted if the series $\boldsymbol{\varphi}^{\mathrm{T}}\boldsymbol{Z}$ is finite (in which case the obtained EL equations are only approximate). For infinite series, the term-wise differentiation should be somehow established, e.g. as is done in AP17. Term-wise differentiation of poorly convergent series may lead to erroneous results; cf. the discussion by (Godin, 1998), Appendix B, in relation with coupled-mode systems for hydroacoustic problems.

Combining Eqs. (3.5) and (3.8), the variation $\delta_{\varphi_m}\tilde{\mathcal{S}}[\eta,\boldsymbol{\varphi};\delta\varphi_m]$, Eq. (3.5), is finally written as

$$\begin{aligned}
&\delta_{\varphi_m}\tilde{\mathcal{S}}[\eta,\boldsymbol{\varphi};\delta\varphi_m] = \\
&= \int_{t_0}^{t_1}\int_X \left\{ \left(-\partial_t\eta + \boldsymbol{N}_\eta\cdot\left[\nabla(\boldsymbol{\varphi}^{\mathrm{T}}\boldsymbol{Z})\right]_{z=\eta}\right)\left[Z_m\right]_{z=\eta} + \sum_n L_{mn}[\eta,h]\,\varphi_n \right\} \delta\varphi_m\, d\boldsymbol{x}\, dt \\
&\qquad + \int_{t_0}^{t_1}\int_{\mathbb{R}} \left(\sum_n T_{mn}\,\varphi_n - g_m\right)\left[\delta\varphi_m\right]_{x_1=a} dx_2\, dt,
\end{aligned} \tag{3.9}$$

where $g_m = \int_{-h_a}^{\eta_a} V_a \left[Z_m\right]_{x_1=a} dz$.

We shall now turn to the calculation of the variation $\delta_\eta\tilde{\mathcal{S}}[\eta,\boldsymbol{\varphi};\delta\eta]$. In view of (3.4a), we replace $\Phi$ and $\delta\Phi$ in Eq. (2.7b), by $\boldsymbol{\varphi}^{\mathrm{T}}\boldsymbol{Z}(\eta)$ and $(\boldsymbol{\varphi}^{\mathrm{T}}\partial_\eta\boldsymbol{Z}(\eta))\,\delta\eta$, respectively. Performing similar calculations as above, and taking into account that $\left[\delta\eta\right]_{x_1=a} = 0$, we obtain

$$\begin{aligned}
&\delta_\eta\tilde{\mathcal{S}}[\eta,\boldsymbol{\varphi};\delta\eta] = \\
&= \int_{t_0}^{t_1}\int_X \left\{ \left[\partial_t(\boldsymbol{\varphi}^{\mathrm{T}}\boldsymbol{Z})\right]_{z=\eta} + \frac{1}{2}\left[\nabla(\boldsymbol{\varphi}^{\mathrm{T}}\boldsymbol{Z})\right]_{z=\eta}^2 + g\eta - \sum_m\left(\sum_n \ell_{mn}[\eta,h]\varphi_n\right)\varphi_m \right. \\
&\qquad \left. + \left(-\partial_t\eta + \boldsymbol{N}_\eta\cdot\left[\nabla(\boldsymbol{\varphi}^{\mathrm{T}}\boldsymbol{Z})\right]_{z=\eta}\right)\left[\boldsymbol{\varphi}^{\mathrm{T}}\partial_\eta\boldsymbol{Z}\right]_{z=\eta} + \frac{p_{\mathrm{surf}}}{\rho} \right\} \delta\eta\, d\boldsymbol{x}\, dt,
\end{aligned} \tag{3.10}$$

where

$$\sum_n \ell_{mn}\,\varphi_n = \int_{-h}^{\eta} \Delta(\boldsymbol{\varphi}^{\mathrm{T}}\boldsymbol{Z})\,\partial_\eta Z_m\, dz - \boldsymbol{N}_h\cdot\left[\nabla(\boldsymbol{\varphi}^{\mathrm{T}}\boldsymbol{Z})\,\partial_\eta Z_m\right]_{z=-h}, \tag{3.11}$$

and $\ell_{mn}[\eta,h]$ are linear differential operators given by

$$\ell_{mn}[\eta,h]\,\bullet = a_{mn}(\eta,h)\nabla_{\boldsymbol{x}}^2\,\bullet + \boldsymbol{b}_{mn}(\eta,h)\cdot\nabla_{\boldsymbol{x}}\,\bullet + c_{mn}(\eta,h)\,\bullet, \tag{3.12}$$

with

$$a_{mn} = \int_{-h}^{\eta} Z_n \, \partial_\eta Z_m \, dz\,, \tag{3.13a}$$

$$\boldsymbol{b}_{mn} = (b_{mn}^{(1)}, b_{mn}^{(2)}) = 2\int_{-h}^{\eta} (\nabla_{\boldsymbol{x}} Z_n)\,\partial_\eta Z_m \, dz + \nabla_{\boldsymbol{x}} h \left[ Z_n \partial_\eta Z_m \right]_{z=-h}, \tag{3.13b}$$

$$c_{mn} = \int_{-h}^{\eta} (\nabla_{\boldsymbol{x}}^2 Z_n + \partial_z^2 Z_n)\,\partial_\eta Z_m \, dz - \boldsymbol{N}_h \cdot \left[ (\nabla_{\boldsymbol{x}} Z_n, \partial_z Z_n)\,\partial_\eta Z_m \right]_{z=-h}. \tag{3.13c}$$

Substituting the functional derivatives $\delta_\eta \tilde{\mathcal{S}}[\eta, \boldsymbol{\varphi}; \delta\eta]$ and $\delta_{\varphi_m} \tilde{\mathcal{S}}[\eta, \boldsymbol{\varphi}; \delta\varphi_m]$ from Eqs. (3.9) and (3.10), into the variational equation (3.3), and using the arbitrariness of the variations $(\delta\eta, \delta\boldsymbol{\varphi})$, yields the following EL equations

$$\delta\eta: \quad \left[\partial_t(\boldsymbol{\varphi}^{\mathrm{T}}\boldsymbol{Z})\right]_{z=\eta} + g\eta + \frac{1}{2}\left[\nabla(\boldsymbol{\varphi}^{\mathrm{T}}\boldsymbol{Z})\right]_{z=\eta}^2 - \sum_m \left(\sum_n \ell_{mn}[\eta, h]\varphi_n\right)\varphi_m$$
$$+\left(-\partial_t\eta + \boldsymbol{N}_\eta \cdot \left[\nabla(\boldsymbol{\varphi}^{\mathrm{T}}\boldsymbol{Z})\right]_{z=\eta}\right)\left(\boldsymbol{\varphi}^{\mathrm{T}}\left[\partial_\eta \boldsymbol{Z}\right]_{z=\eta}\right) = -\frac{p_{\text{surf}}}{\rho}, \tag{3.14a}$$

$$\delta\varphi_m: \quad \left(\partial_t\eta - \boldsymbol{N}_\eta \cdot \left[\nabla(\boldsymbol{\varphi}^{\mathrm{T}}\boldsymbol{Z})\right]_{z=\eta}\right)\left[Z_m\right]_{z=\eta} + \sum_n L_{mn}[\eta, h]\varphi_n = 0, \tag{3.14b}$$

$$\left[\delta\varphi_m\right]_{x_1=a}: \quad \sum_n T_{mn}\varphi_n = g_m. \tag{3.14c}$$

The system (3.14b) corresponds to the kinematical equations (2.2a-c) (see also Lemma 1, below), while Eq. (3.14a) corresponds to the dynamical (Bernoulli) condition on the free surface, Eq. (2.2d). Eq. (3.14c) corresponds to a Neumann lateral condition on the line $\{x_1 = a\}$; cf. Eq. $(2.3b)$. The above equations must be supplemented by the rest-at-infinity conditions, $\eta, \varphi_n, \nabla_{\boldsymbol{x}}\varphi_n \to 0$ as $|\boldsymbol{x}| \to \infty$, which are essential conditions of the variational formulation.

It is worth noting here that no simplifying assumptions have been explicitly used in the derivation of Eqs. (3.14a,b,c). More specifically, all nonlinearities have been treated "exactly" and they are present in Eqs. (3.14a,b), and all horizontal derivatives on the seabed and the free surface have been kept in the formulation; see the formulae for coefficients $\boldsymbol{B}_{mn}, C_{mn}$, Eqs. (3.7b,c), and $\boldsymbol{b}_{mn}, c_{mn}$, Eqs. (3.13b,c). Nevertheless, Eqs. (3.14) cannot be exact if the series expansion $\Phi = \boldsymbol{\varphi}^{\mathrm{T}}\boldsymbol{Z}$, underlying their derivation, is not exact. Thus, as already mentioned in the beginning of the present section, Eqs. (3.14) can be used either as a means for deriving simplified model equations, based on approximate (finite) series representations of the wave potential, or as an intermediate step towards a new Hamiltonian formulation of the exact NLIWW problem, based on an exact, convergent, infinite series expansion of the wave potential. The latter direction will be exploited in the next section. In the remaining of the present section, we shall briefly demonstrate how various families of simplified water-wave models can be (re)derived by using Eqs. (3.14).

Linearizing Eqs. (3.14), and prescribing the vertical functions $Z_n$ to be the eigenfunctions of the linear water-wave problem, supported on the interval $[-h(\boldsymbol{x}), 0]$, we obtain the time-dependent version of multimodal mild-slope equations, first derived, by different methods, by Massel (1993)and Porter & Staziker (1995). Although these equations contain terms of order $O(\nabla_x h, \nabla_x^2 h)$, in contrast with the standard mild-slope equation of Berkhof, they are still considered "mild-slope" due to poor convergence of the employed eigenfunctions expansions near the seabed; see relevant discussions in Athanassoulis & Belibassakis (1999), (Chamberlain & Porter, 2006). Enriching the aforementioned eigenfunctions expansions by an appropriate additional function, the linearized version of Eqs. (3.14) leads to the CCMS first derived in Athanassoulis & Belibassakis (1999) and developed in various directions by the same authors; see (Belibassakis, Athanassoulis, & Gerostathis, 2001), (Athanassoulis & Belibassakis, 2007), (Belibassakis & Athanassoulis, 2011) (Gerostathis, Belibassakis, & Athanassoulis, 2008), (Gerostathis, Belibassakis, & Athanassoulis, 2016).

Further, considering Eqs. (3.14) without any formal simplification concerning nonlinearity and choose the vertical functions to be independent of $\eta$, $Z_n = Z_n(z; h(\boldsymbol{x}))$, we obtain the nonlinear, multimodal, mild-slope equation proposed by (Isobe & Abohadima, 1998). These authors show that further simplifications of their equations lead to the standard mild-slope equation, the shallow-water equations, and the Peregrine's form of Boussinesq equations. Closing this section, we show how the Boussinesq-type equations derived by (Klopman et al., 2010) (referred to in the following as KVGD10) are obtained as a special case of Eqs. (3.14). The vertical functions used in KVGD10 have the form

$$Z_0 = 1, \quad Z_n = Z_n(z; \eta, h), \quad \left[ Z_n \right]_{z=\eta} = 0, \quad n \geq 1.$$

Thus, $\partial_t Z_0 = 0$, $\nabla_{\boldsymbol{x}} Z_0 = 0$, $\partial_t Z_n = \partial_\eta Z_n \, \partial_t \eta$, $\nabla_{\boldsymbol{x}} Z_n = \partial_h Z_n \, \nabla_{\boldsymbol{x}} h + \partial_\eta Z_n \, \nabla_{\boldsymbol{x}} \eta$. Elaborating our coefficients $A_{mn}$, $\boldsymbol{B}_{mn}$, $C_{mn}$ and $a_{mn}$, $\boldsymbol{b}_{mn}$, $c_{mn}$ in conformity with Klopman's simplification ($\nabla_{\boldsymbol{x}} h \approx 0$), and performing appropriate algebraic manipulations, we find that our Eq. (3.14a) reduces to Eq. (4.3b) of KVGD10, the first of our Eqs. (3.14b) gives Eq. (4.3a) of KVGD10, and the remaining of our Eqs. (3.14b) become identical with Eqs. (4.3c) of KVGD10.

## 4. Hamiltonian Equations based on an exact representation of the wave potential

### 4.1 *Exact representation of the wave potential*

It has been recently proved in AP17, that any smooth enough function $\Phi(\boldsymbol{x}, z, t)$, defined on the strip-like domain $\overline{D_h^\eta}(X, t)$ (the overbar denotes topological closure of the domain), can be represented by a rapidly convergent series of the form

$$\Phi(\boldsymbol{x},z,t) = \varphi_{-2}(\boldsymbol{x},t)\, Z_{-2}(z;\eta,h) + \varphi_{-1}(\boldsymbol{x},t)\, Z_{-1}(z;\eta,h) + \sum_{n=0}^{\infty} \varphi_n(\boldsymbol{x},t)\, Z_n(z;\eta,h), \tag{4.1}$$

where the sequence $\{Z_n(z;\eta,h),\, n \geq 0\}$ is an $L^2$ – basis in $[-h,\eta] = [-h(\boldsymbol{x}), \eta(\boldsymbol{x},t)]$, and $Z_{-2}(z;\eta,h)$, $Z_{-1}(z;\eta,h)$ are two appropriate additional vertical functions, first in-troduced heuristically by Athanassoulis & Belibassakis (2000). A convenient choice of the vertical functions $Z_{-2}$, $Z_{-1}$, $\{Z_n, n \geq 0\}$ is:

$$Z_{-2}(z;\eta,h) = \frac{\mu_0 h_0 + 1}{2 h_0} \frac{(z+h)^2}{\eta + h} - \frac{\mu_0 h_0 + 1}{2 h_0}(\eta + h) + 1, \qquad (4.2a)$$

$$Z_{-1}(z;\eta,h) = \frac{\mu_0 h_0 - 1}{2 h_0} \frac{(z+h)^2}{\eta + h} + \frac{1}{h_0}(z+h) - \frac{\mu_0 h_0 + 1}{2 h_0}(\eta + h) + 1, \quad (4.2b)$$

$$Z_0(z;\eta,h) = \frac{\cosh[k_0(z+h)]}{\cosh[k_0(\eta+h)]}, \qquad Z_n(z;h,\eta) = \frac{\cos[k_n(z+h)]}{\cos[k_n(\eta+h)]}. \qquad (4.2c,d)$$

In Eqs. (4.2), $k_n = k_n(\boldsymbol{x},t)$, $n \geq 0$ are functions implicitly defined by the (local) transcendental equations

$$\mu_0 - k_0 \tanh[k_0(\eta+h)] = 0, \qquad \mu_0 + k_n \tan[k_n(\eta+h)] = 0, \qquad (4.3a,b)$$

where $\mu_0 > 0$ is an arbitrary constant, and $h_0$ (reference depth) is introduced for dimensional consistency. Note that all vertical functions $Z_n$ in Eqs. (4.2) have been chosen so that to satisfy the normalization condition

$$[Z_n]_{z=\eta} = 1, \quad n \geq -2, \qquad (4.4)$$

which implies $\sum_{n=-2}^{\infty} \varphi_n = [\Phi]_{z=\eta}$.

More specifically, in AP17 it was shown that, if the boundary functions $h(\boldsymbol{x})$, $\eta(\boldsymbol{x},t)$ and the field $\Phi(\boldsymbol{x},z,t)$ are smooth enough, then $\varphi_n(\boldsymbol{x},t)$, $\partial_{x_i}\varphi_n(\boldsymbol{x},t)$ and $\partial^2_{x_i}\varphi_n(\boldsymbol{x},t)$ are all of order $O(n^{-4})$, uniformly in $(\boldsymbol{x},t)$, and the series expansion (4.1) can be term-wise differentiated two times in $\overline{D_h^\eta}(X,t)$. These properties will be frequently invoked in the sequel, where the results of Section 3 will be given a rigorous and more convenient form, revealing the Hamiltonian structure of the NLIWW problem, along with a new, efficient form of the DtN operator.

**Remark:** All smoothness requirements (not explicitly stated above) hold true for the wave potential $\Phi(\boldsymbol{x},z,t)$, since it is an harmonic function and, thus, it is infinitely differentiable in the open domain $D_h^\eta(X)$. Accordingly, for sufficiently smooth boundary functions $h(\boldsymbol{x})$, $\eta(\boldsymbol{x},t)$ all convergence properties stated above are valid, and they are legitimately invoked in the analysis that follows.

### 4.2 *EL equations corresponding to the representation* (4.1)

The convergence properties of the series expansion (4.1) ensures that the transformation $(\eta,\Phi) \leftrightarrow (\eta,\{\varphi_n(\boldsymbol{x},t)\}_{n=-2}^{\infty}) \equiv (\eta,\boldsymbol{\varphi})$ is one-to-one, invertible and smooth. This implies that the critical "points" of the action functionals $\mathcal{S}[\eta,\Phi]$ and $\tilde{\mathcal{S}}[\eta,\boldsymbol{\varphi}]$ are essentially the same, in the sense that

$$\delta\mathcal{S}[\eta,\Phi;\delta\eta,\delta\Phi] = 0 \quad \Leftrightarrow \quad \delta\tilde{\mathcal{S}}[\eta,\boldsymbol{\varphi};\delta\eta,\delta\boldsymbol{\varphi}] = 0.$$

Therefore, in this case, Eqs. (3.14) are not just an approximate model of the problem (2.2), but they provide an equivalent reformulation of it. Further, in this case, Eqs. (3.14) may be given a much simpler form, without sacrificing their exactness.

Let us first focus on the system of Eqs. (3.14b). Using the normalization condition (4.4), Eqs. (3.14b) take the form

$$\partial_t \eta - \boldsymbol{N}_\eta \cdot \left[ \nabla(\boldsymbol{\varphi}^{\mathrm{T}} \boldsymbol{Z}) \right]_{z=\eta} + \sum_{n=-2}^{\infty} L_{mn}[\eta,h]\varphi_n = 0, \; m \geq -2, \quad \boldsymbol{x} \in X. \tag{4.5}$$

Note that, according to Theorem 2 of AP17, the infinite series appearing in Eqs. (4.5) converge throughout $\overline{D_h^\eta}(X,t)$. In the following lemma, we elaborate further on the system (4.5), obtaining an equivalent decomposition of it, consisting of one evolution equation and an infinite system of time-independent equations.

**Lemma 1:** *Assume that* $\boldsymbol{\varphi} = \{\varphi_n(\boldsymbol{x};t)\}_{n=-2}^{\infty}$ *satisfies the system* (4.5) *at every* $t \in [t_0, t_1]$. *Then, the function* $\Phi = \boldsymbol{\varphi}^{\mathrm{T}} \boldsymbol{Z}$, *where* $\boldsymbol{Z} = \{Z_n(\boldsymbol{x},z,t)\}_{n=-2}^{\infty}$ *are given by* Eqs. (4.2), *satisfies* Eqs. (2.2a) *and* (2.2b), *that is*

$$\Delta(\boldsymbol{\varphi}^{\mathrm{T}} \boldsymbol{Z}) = 0, \quad \boldsymbol{N}_h \cdot \left[ \nabla(\boldsymbol{\varphi}^{\mathrm{T}} \boldsymbol{Z}) \right]_{z=-h} = 0. \tag{4.6}$$

*Further, at every* $t \in [t_0, t_1]$, *we have that*

$$\partial_t \eta - \boldsymbol{N}_\eta \cdot \left[ \nabla(\boldsymbol{\varphi}^{\mathrm{T}} \boldsymbol{Z}) \right]_{z=\eta} = 0, \quad \boldsymbol{x} \in X, \tag{4.7a}$$

$$\sum_{n=-2}^{\infty} L_{mn}[\eta,h]\varphi_n = 0, \quad \boldsymbol{x} \in X, \quad m \geq -2. \tag{4.7b}$$

*Proof:* Let us fix an $m = m_*$. Subtracting the $m_*^{\mathrm{th}}$ equation of (4.5) from the others, and using identity (3.8a), we find that $\boldsymbol{\varphi} = \{\varphi_n(\boldsymbol{x};t)\}_{n=-2}^{\infty}$ satisfies

$$\partial_t \eta - \boldsymbol{N}_\eta \cdot \left[ \nabla(\boldsymbol{\varphi}^{\mathrm{T}} \boldsymbol{Z}) \right]_{z=\eta} + \int_{-h}^{\eta} \Delta(\boldsymbol{\varphi}^{\mathrm{T}} \boldsymbol{Z}) Z_{m_*} \, dz - \boldsymbol{N}_h \cdot \left[ \nabla(\boldsymbol{\varphi}^{\mathrm{T}} \boldsymbol{Z}) Z_{m_*} \right]_{z=-h} = 0, \tag{4.8a}$$

$$\int_{-h}^{\eta} \Delta(\boldsymbol{\varphi}^{\mathrm{T}} \boldsymbol{Z})(Z_m - Z_{m_*}) \, dz - \boldsymbol{N}_h \cdot \left[ \nabla(\boldsymbol{\varphi}^{\mathrm{T}} \boldsymbol{Z})(Z_m - Z_{m_*}) \right]_{z=-h} = 0, \quad m \geq -2. \tag{4.8b}$$

We shall first show that Eqs. (4.8b) imply Eqs. (4.6). For this purpose, we consider a smooth function $\delta\Psi$, defined on the closed domain $\overline{D_h^\eta}(X,t)$, being arbitrary in the open domain $D_h^\eta(X,t)$ and on the seabed boundary, and satisfying the condition $[\delta\Psi]_{z=\eta} = 0$ on the free surface. By Theorem 1 of AP17, there exists an admissible sequence $\delta\boldsymbol{\psi} = \{\delta\psi_m(\boldsymbol{x};t)\}_{m=-2}^{\infty}$ such that $\delta\Psi = \sum_{m=-2}^{\infty} \delta\psi_m Z_m$. Multiplying the $m^{th}$ of Eqs. (4.8b) by $\delta\psi_m$, integrating over $X$, and summing over $m$, we obtain:

$$\int_X \int_{-h}^{\eta} \Delta(\boldsymbol{\varphi}^{\mathrm{T}} \boldsymbol{Z}) \sum_{m=-2}^{\infty} \delta\psi_m (Z_m - Z_{m_*})\, dz\, d\boldsymbol{x} \\ - \int_X \boldsymbol{N}_h \cdot \left[ \nabla(\boldsymbol{\varphi}^{\mathrm{T}} \boldsymbol{Z}) \right]_{z=-h} \sum_{m=-2}^{\infty} \delta\psi_m \left[ Z_m - Z_{m_*} \right]_{z=-h} d\boldsymbol{x} = 0. \tag{4.9}$$

Consider now the function $\delta\tilde{\Psi} = \sum_{m=-2}^{\infty} \delta\psi_m (Z_m - Z_{m_*})$. Since

$$\delta\tilde{\Psi} = \sum_{m=-2}^{\infty} \delta\psi_m Z_m - Z_{m_*} \sum_{m=-2}^{\infty} \delta\psi_m = \delta\Psi - Z_{m_*} [\delta\Psi]_{z=\eta},$$

and $[\delta\Psi]_{z=\eta} = 0$ by construction, we conclude that $\delta\tilde{\Psi} = \delta\Psi$. Thus, Eq. (4.9) takes the form

$$\int_X \int_{-h}^{\eta} \Delta(\boldsymbol{\varphi}^{\mathrm{T}} \boldsymbol{Z}) \delta\Psi\, dz\, d\boldsymbol{x} - \int_X \boldsymbol{N}_h \cdot \left[ \nabla(\boldsymbol{\varphi}^{\mathrm{T}} \boldsymbol{Z}) \right]_{z=-h} [\delta\Psi]_{z=-h} d\boldsymbol{x} = 0. \tag{4.10}$$

Since $\delta\Psi$ is arbitrary in the open domain $D_h^{\eta}(X,t)$ and on the seabed boundary, by using the standard arguments of the calculus of variations, we obtain Eqs. (4.6). This result, in conjunction with the identity (3.8a), leads to $\sum_{n=-2}^{\infty} L_{mn}[\eta,h]\varphi_n = 0$ for any $m \geq -2$, which proves Eqs. (4.7b). Then, Eq. (4.5) reduces to (4.7a). Eq. (4.7a) can also be obtained by Eq. (4.8a), observing that the two last terms vanish identically, in view of the already proven Eqs. (4.6).

Lemma 1 permits us to drastically simplify Eq. (3.14a), as well. Observe first that Eqs. (4.6) and the identity (3.11) imply that $\sum_{n=-2}^{\infty} \ell_{mn}[\eta,h]\,\varphi_n = 0$. Thus, using also Eq. (4.7a), we see that Eq. (3.14a) takes the much simpler form

$$\left[ \partial_t(\boldsymbol{\varphi}^{\mathrm{T}} \boldsymbol{Z}) \right]_{z=\eta} + g\eta + \frac{1}{2} \left[ \nabla(\boldsymbol{\varphi}^{\mathrm{T}} \boldsymbol{Z}) \right]_{z=\eta}^2 = -\frac{p_{\mathrm{surf}}}{\rho}. \tag{4.11}$$

Eqs. (4.7) and (4.11) constitute an exact reformulation of the NLIWW problem, in terms of the free surface elevation $\eta$ and the modal amplitudes $\boldsymbol{\varphi} = \{\varphi_n(\boldsymbol{x};t)\}_{n=-2}^{\infty}$. One can easily identify the content of Eqs. (4.7) and (4.11) with the kinematic equations (2.2a-c) and the dynamic condition (2.2d), respectively. Despite their exactness, Eqs. (4.7) and (4.11) contain the infinite series $\boldsymbol{\varphi}^{\mathrm{T}} \boldsymbol{Z} = \sum \varphi_n Z_n$, making them inconvenient for numerical treatment. This unpleasant feature is remedied in the following subsection.

### 4.3 *Derivation of Hamiltonian equations*

We introduce the free-surface potential $\psi(\boldsymbol{x},t) = \boldsymbol{\varphi}^{\mathrm{T}} [\boldsymbol{Z}]_{z=\eta} = \sum_{n=-2}^{\infty} \varphi_n$, and proceed with the following

**Lemma 2:** *For each $t \in [t_0, t_1]$, the infinite system*

$$\sum_{n=-2}^{\infty} L_{mn}[\eta,h]\varphi_n = 0, \qquad \boldsymbol{x}\in X, \quad m\geq -2, \tag{4.12a}$$

$$\sum_{n=-2}^{\infty} \varphi_n = \psi, \qquad \boldsymbol{x}\in X, \tag{4.12b}$$

$$\sum_{n=-2}^{\infty} T_{mn}\varphi_n = g_m, \quad x_2\in\mathbb{R}, \quad m\geq -2, \qquad \eta, \varphi_n, \nabla_{\boldsymbol{x}}\varphi_n \to 0 \ \text{ as } |\boldsymbol{x}|\to\infty \tag{4.12c,d}$$

*is equivalent to the boundary value problem*

$$\Delta\Phi = 0, \quad \text{on } D_h^{\eta}(X,t), \tag{4.13a}$$

$$\boldsymbol{N}_h\cdot[\nabla\Phi]_{z=-h} = 0, \quad [\Phi]_{z=\eta} = \psi, \qquad \boldsymbol{x}\in X, \tag{4.13b,c}$$

$$\left[\partial_{x_1}\Phi\right]_{x_1=a} = V_a, \quad x_2\in\mathbb{R}, \qquad \eta, |\nabla\Phi| \to 0 \ \text{ as } |(\boldsymbol{x},z)|\to\infty. \tag{4.13d,e}$$

***Proof***: Let $\Phi$ be the (unique) classical solution of problem (4.13), and $\boldsymbol{\varphi}^{\mathrm{T}}\boldsymbol{Z}$ its modal expansion (Theorem 1 of AP17). Then, Eqs. (4.13a,b), in conjunction with the identity (3.8a) yield Eqs. (4.12a), while Eq. (4.13c) in conjunction with the normalization condition (4.4) gives directly Eq. (4.12b), since $[\Phi]_{z=\eta} = \boldsymbol{\varphi}^{\mathrm{T}}[\boldsymbol{Z}]_{z=\eta} = \sum_{n=-2}^{\infty}\varphi_n$. In order to obtain the boundary condition (4.12c) on the excitation boundary, we multiply the first of Eqs. (4.13d) by $Z_m$, integrate over $[-h_a, \eta_a]$ and recall that $\Phi = \boldsymbol{\varphi}^{\mathrm{T}}\boldsymbol{Z}$ and $g_m = \int_{-h_a}^{\eta_a} V_a [Z_m]_{x_1=a}\, dz$. Finally, Eq. (4.13e) gives $|\nabla\Phi|^2 = |\nabla_{\boldsymbol{x}}\boldsymbol{\varphi}\boldsymbol{Z} + \boldsymbol{\varphi}\nabla_{\boldsymbol{x}}\boldsymbol{Z}|^2 + |\boldsymbol{\varphi}\,\partial_z\boldsymbol{Z}|^2 \to 0$ which implies Eq. (4.12d), since $\boldsymbol{Z}$ and $\partial_z\boldsymbol{Z}$ do not vanish at infinity.

Now, suppose that the (admissible) sequence $\boldsymbol{\varphi} = \{\varphi_n\}_{n=-2}^{\infty}$ satisfies Eqs. (4.12) and consider the field reconstructed by $\Phi = \boldsymbol{\varphi}^{\mathrm{T}}\boldsymbol{Z}$. We see at once that Eqs. (4.12b) implies Eq. (4.13c). In order to prove (4.13a,b), take an arbitrary admissible function $\delta\Psi$, defined on the closed domain $\overline{D_h^{\eta}(X,t)}$, and consider the corresponding admissible sequence $\delta\boldsymbol{\psi} = \{\delta\psi_m(\boldsymbol{x};t)\}_{m=-2}^{\infty}$ such that $\delta\Psi = \sum_{m=-2}^{\infty}\delta\psi_m Z_m$. Multiplying the $m^{th}$ Eq. (4.12a) by $\delta\psi_m$, integrating over $X$, and summing over $m$, we obtain an equation like Eq. (4.10). Then, we continue by using variational arguments, as in Lemma 1. The lateral excitation condition (4.13d) is obtained similarly, by using (4.12c). Finally, it is easy to verify that $|\nabla\Phi|\to 0$ as $|(\boldsymbol{x},z)|\to\infty$, when Eqs. (4.12d) hold. □

The system of Eqs. (4.12), called the *substrate* system, determines the sequence of modal amplitudes $\boldsymbol{\varphi} = \{\varphi_n\}_{n=-2}^{\infty}$ in terms of the surface fields $(\eta,\psi)$, at any specific time $t$. We emphasize this fact by writing $\varphi_n = \mathcal{F}_n[\eta,h]\psi$ for the solution of Eqs. (4.12). It remains to re-write the evolution equations (4.7a) and (4.11) as evolution equations on $(\eta,\psi)$. To this end, we need to express the traces of the temporal and spatial derivatives of the series $\boldsymbol{\varphi}^{\mathrm{T}}\boldsymbol{Z}$ on the free surface in terms of $\eta$ and $\psi$.

**Lemma 3:** *For vertical functions* $Z_n = Z_n(z\,;\eta(\boldsymbol{x},t),h(\boldsymbol{x}))$ *given by* Eqs. (4.2a), *the following equations hold true*:

$$\left[\partial_t(\boldsymbol{\varphi}^{\mathrm{T}}\boldsymbol{Z})\right]_{z=\eta} = \partial_t\psi - \boldsymbol{\varphi}^{\mathrm{T}}\left[\partial_z\boldsymbol{Z}\right]_{z=\eta}\partial_t\eta, \tag{4.14a}$$

$$\left[\nabla_{\boldsymbol{x}}(\boldsymbol{\varphi}^{\mathrm{T}}\boldsymbol{Z})\right]_{z=\eta} = \nabla_{\boldsymbol{x}}\psi - \boldsymbol{\varphi}^{\mathrm{T}}\left[\partial_z\boldsymbol{Z}\right]_{z=\eta}\nabla_{\boldsymbol{x}}\eta, \tag{4.14b}$$

$$\left[\partial_z(\boldsymbol{\varphi}^{\mathrm{T}}\boldsymbol{Z})\right]_{z=\eta} = \boldsymbol{\varphi}^{\mathrm{T}}\left[\partial_z\boldsymbol{Z}\right]_{z=\eta} = h_0^{-1}\varphi_{-2} + \mu_0\psi\,. \tag{4.14c}$$

***Proof***: To prove Eq. (4.14a), calculate first $\partial_t(\boldsymbol{\varphi}^{\mathrm{T}}\boldsymbol{Z})$ at an interior point $-h<z<\eta$, $\partial_t(\boldsymbol{\varphi}^{\mathrm{T}}\boldsymbol{Z}) = (\partial_t\boldsymbol{\varphi}^{\mathrm{T}})\boldsymbol{Z} + \boldsymbol{\varphi}^{\mathrm{T}}(\partial_\eta\boldsymbol{Z})\,\partial_t\eta$, and, then, take the limit as $z\to\eta$, obtaining

$$\left[\partial_t(\boldsymbol{\varphi}^{\mathrm{T}}\boldsymbol{Z})\right]_{z=\eta} = \partial_t\boldsymbol{\varphi}^{\mathrm{T}}\left[\boldsymbol{Z}\right]_{z=\eta} + \boldsymbol{\varphi}^{\mathrm{T}}\left[\partial_\eta\boldsymbol{Z}\right]_{z=\eta}\partial_t\eta\,.$$

On the other hand, since $\left[\boldsymbol{Z}\right]_{z=\eta} = \boldsymbol{Z}(z=\eta(\boldsymbol{x},t)\,;\eta(\boldsymbol{x},t),h(\boldsymbol{x}))$, we have

$$\partial_t\left[\boldsymbol{\varphi}^{\mathrm{T}}\boldsymbol{Z}\right]_{z=\eta} = \partial_t\boldsymbol{\varphi}^{\mathrm{T}}\left[\boldsymbol{Z}\right]_{z=\eta} + \boldsymbol{\varphi}^{\mathrm{T}}\left(\left[\partial_z\boldsymbol{Z}\right]_{z=\eta} + \left[\partial_\eta\boldsymbol{Z}\right]_{z=\eta}\right)\partial_t\eta\,.$$

Subtracting the latter equation from the former and taking into account that $\left[\boldsymbol{\varphi}^{\mathrm{T}}\boldsymbol{Z}\right]_{z=\eta}=\psi$, we obtain Eq. (4.14a). Eq. (4.14b) is proved similarly, while the first equality in Eq. (4.14c) is obvious since $\boldsymbol{\varphi}$ is $z$ independent. To prove the second equality of Eq. (4.14c), use is made of the equations $\left[Z_n\right]_{z=\eta} = 1$, $\left[\partial_z Z_{-2}\right]_{z=\eta} = \mu_0 + h_0^{-1}$, and $\left[\partial_z Z_n\right]_{z=\eta} = \mu_0$, $n\geq -1$, which hold true for the specific choice of vertical functions, Eqs. (4.2a). □

Substituing Eqs. (4.14) into Eqs. (4.7a) and (4.11), we arrive at the following

**Theorem 1:** *The NLIWW problem is equivalent to the following system of two nonlinear and nonlocal evolution equations*

$$\partial_t\eta = -(\nabla_{\boldsymbol{x}}\eta)\cdot(\nabla_{\boldsymbol{x}}\psi) + (|\nabla_{\boldsymbol{x}}\eta|^2+1)\left(h_0^{-1}\mathcal{F}_{-2}[\eta,h]\psi + \mu_0\psi\right), \tag{4.15a}$$

$$\partial_t\psi = -g\eta - \frac{1}{2}(\nabla_{\boldsymbol{x}}\psi)^2 + \frac{1}{2}(|\nabla_{\boldsymbol{x}}\eta|^2+1)\left(h_0^{-1}\mathcal{F}_{-2}[\eta,h]\psi + \mu_0\psi\right)^2 - \frac{p_{\mathrm{surf}}}{\rho}, \tag{4.15b}$$

*where* $\mathcal{F}_{-2}[\eta,h]\psi = \varphi_{-2}(\boldsymbol{x},t)$ *is the first element of the modal sequence* $\{\varphi_n(\boldsymbol{x};t)\}_{n=-2}^{\infty}$ *obtained by solving the modal substrate problem* (4.12).

The above result has been first announced in (Athanassoulis & Papoutsellis, 2015), without proof. Closing this section, we establish the connection of the above new formulation with the usual Hamiltonian formulation of water waves, introduced by (Zakharov, 1968), and expressed in terms of the DtN operator by (W. Craig & Sulem, 1993). Recall first that the DtN operator is defined by $G[\eta,h]\psi = \boldsymbol{N}_\eta \cdot [\nabla\Phi]_{z=\eta}$, where $\Phi$ solves Eqs. (4.13). Comparing Eq. (4.15a) with the kinematic condition (2.2c), written in terms of $G[\eta,h]\psi$, we easily conclude that

$$G[\eta,h]\psi = -(\nabla_{\mathbf{x}}\eta)\cdot(\nabla_{\mathbf{x}}\psi) + (|\nabla_{\mathbf{x}}\eta|^2+1)\left(h_0^{-1}\mathcal{F}_{-2}[\eta,h]\psi + \mu_0\psi\right). \qquad (4.16)$$

In the existing literature, the DtN operator is defined under the assumptions of a flat horizontal seabed and spatial periodicity, and treated by perturbative techniques (see e.g. ( Craig & Sulem, 1993), (Bateman et al., 2001), (Nicholls, 2007)). Although this approach is mathematically nice and numerically effective under the aforementioned assumptions, its extension to general bathymetry is quite complicated and, to the best of our knowledge, has been achieved only for the case where the seabed is a small perturbation of the horizontal plane ((Smith, 1998), (Craig et al., 2005),(Gouin et al., 2016)). The main advantage of Eq. (4.16) is that it is not restricted by any mild-slope assumptions on the physical boundaries of the fluid domain, or horizontal periodicity. The substrate problem (4.12) communicates with the Hamiltonian Eqs. (4.15) only by means of the free surface mode $\varphi_{-2}$, which exhibits a superconvergence with respect to the total number of modes kept in the truncated series expansion (4.1), even for strongly deformed domains; see AP17, Sec. 4(b) and Appendix E. In addition, the solution of (4.12) provides the whole sequence $\boldsymbol{\varphi} = \{\mathcal{F}_n[\eta, h]\psi\}_{n=-2}^{\infty}$, that reconstructs the instantaneous velocity potential $\Phi$ throughout the whole fluid domain, in the form of a rapidly convergent series. Thus, it permits us to easily calculate wave kinematics and pressure fields in complicated domains and for strongly nonlinear waves. This feature will be exploited in a subsequent work.

## 5. Numerical implementation and validation with experiments

### 5.1. *Numerical scheme*

In order to evolve Eqs. (4.15) in time, the knowledge of the nonlocal operator $\mathcal{F}_{-2}[\eta,h]\psi$ at every time $t$ is required. An approximation of $\mathcal{F}_{-2}[\eta,h]\psi$, at time $t$, is computed by solving the truncated substrate system (4.12), retaining a total number of modes $N_{\text{tot}}$ in the series expansion (4.1), $\Phi = \sum_{n=-2}^{M}\varphi_n Z_n$ ($N_{\text{tot}} = M+3$). In the case of one horizontal dimension, the truncated system is composed by the first $N_{\text{tot}}-1$ differential equations (4.12a), the truncated algebraic constraint (4.12b), $\sum_{n=-2}^{M}\varphi_n = \psi$ and the boundary conditions $\sum_{n=-2}^{M} T_{mn}\varphi_n = g_m$, $m=-2,\ldots,M$, at $x=a$. The semi-infinite horizontal domain is truncated at a point $x=b$, where the conditions $\sum_{n=-2}^{M}\left[A_{mn}\partial_x\varphi_n + (1/2)B_{mn}\varphi_n\right]_{x=b} = 0$, $m=-2,\ldots,M$ are applied, accounting for a reflecting vertical wall. The variable coefficients $A_{mn}$, $B_{mn}$, $C_{mn}$, Eqs. (3.7), needed for the numerical set-up of the substrate system (4.12), are efficiently evaluated using available analytical formulas (Papoutsellis, Athanassoulis, & Charalambopoulos, 2017), expressing them in closed-form, in terms of $\eta$, $h$, the parameters $\mu_0$, $h_0$, and the local eigenvalues $\{k_n(\eta,h)\}_{n=0}^{M}$, computed by solving Eqs. (4.3) via the Newton-Raphson method. Given $\eta(x,t)$, $\psi(x,t)$ and $g_m(t)$, the truncated system is discretized by the finite-difference method with fourth order accuracy on a grid $\{x_i\}_{i=1}^{N_X}$ of uniform spacing $\delta x$. The solution of the resulting square linear system leads to the local values of the sequence $\{\varphi_n(x_i)\}_{n=-2}^{M}$ which furnishes an approximation of $\mathcal{F}_{-2}[\eta,h]\psi = \varphi_{-2}$, corresponding to $N_{\text{tot}}$. This approximation is then used in the evolutionary system (4.15) which is marched at time $t+\delta t$ by the classical, fourth-order, Runge-Kutta method. In order to create incident regular wave conditions, the data for $g_m(t)$ is constructed by using highly-accurate, travelling-wave, periodic solutions (having period, amplitude and depth as in the corresponding experiment) using an appropriate variant of the new Hamiltonian formulation, Eqs. (4.15). The excitation starts form quiescence ($\eta(x,0)=\psi(x,0)=0$ and $g_m(0)=0$) and gradually reaches the desired values, according to a smooth transition (ramp) time function. Stable wave generation and absorption is achieved by employing a standard sponge-layer technique at regions in the vicinity of the boundaries of the computational domain; see e.g. (Zhang, Kennedy, Panda, Dawson, & Westerink, 2014).

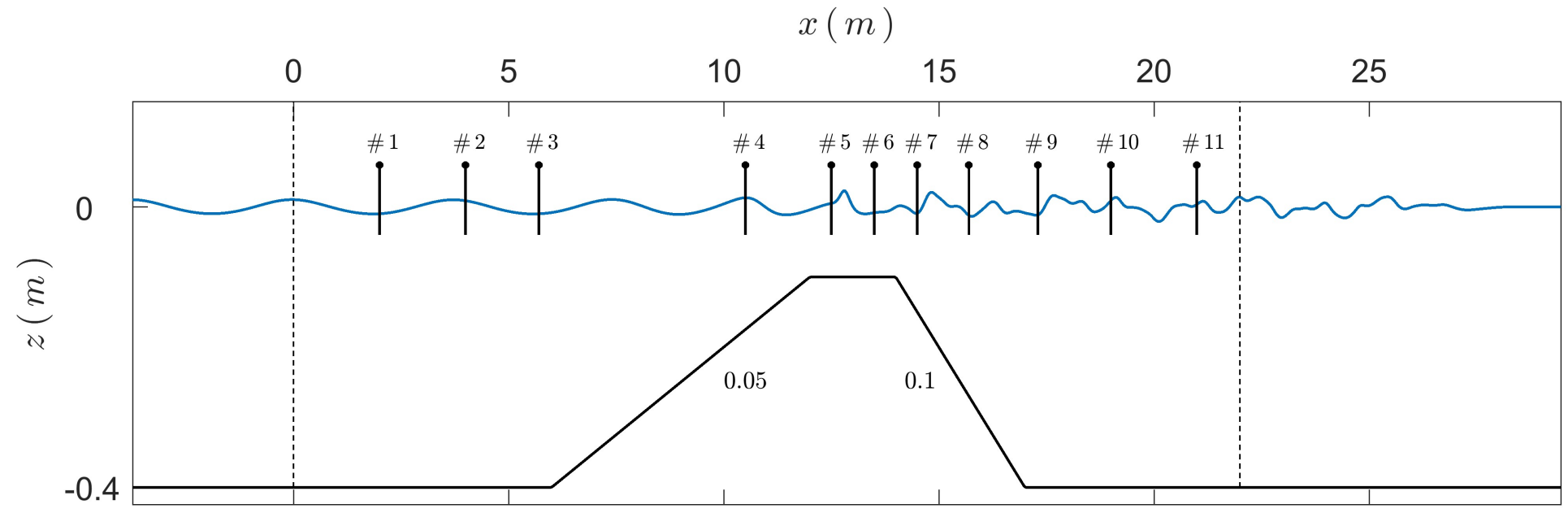

**Figure 2**. Configuration of the numerical wave tank and locations of the measuring stations 1-11 used in (Dingemans, 1994). Dashed lines correspond to the ends of the sponge layers.

### 5.2. *Submerged trapezoid with mild front and back slopes*

The performance of the proposed numerical scheme is first illustrated by simulating the classical experiment of (Beji & Battjes, 1993) and (Dingemans, 1994), who investigated the transformation that an incident long wave undergoes as it propagates in a region with a trapezoidal shoal; figure 2. At the left boundary of the wave-tank, periodic waves of amplitude $H = 0.02\ m$ and period $T = 2$ sec are generated and propagate towards the submerged trapezoidal bar with front and back slopes given respectively by 0.05 and 0.1. As the wave approaches the front side of the bar, the shoaling process is initiated and the wave amplitude increases. When the wave reaches the horizontal side of the bar, it steepens, exhibiting a signifi-cant increase of wave height, just before it is abruptly released towards the deeper region behind the bar, where it is transformed to a rapidly changing dispersive wave pattern. Here, we use the configuration of (Dingemans, 1994), in which the free-surface elevation is recorded at eleven stations distributed along the wave tank (see figure 2). In the simulations, the parameter $\mu_0$ is chosen on the basis of the angular frequency of the incoming field $\omega_{in} = 2\pi / T$, i.e., $\mu_0 = \omega_{in}^2 / g = 0.9855$, and $h_0$ is taken to be the depth of the incidence region, $h_0 = 0.4$ m. The total length of the computational domain (including sponge layers) is 33.2 m, and the spatio-temporal discretization is $\delta x = 0.0468$ m and $\delta t = 0.0202$ sec. Computations obtained by using $N_{\text{tot}} = 7$ modes are compared, in figure 3, with the experimental time series of the free-surface elevation in the time window [35,39] sec, where the wave kinematics is adequately established in the laboratory wave tank. Computations show excellent agreement with the experimental measurements up to Station #11, concerning both the amplitude and the phase of the free surface elevation.

In order to examine how the performance of the present method depends on the number $N_{\text{tot}}$ of modes kept in the truncated modal series expansion, we performed the previous simulation by using $N_{\text{tot}} = 4,5,7,10$ modes, keeping the rest of the numerical parameters unchanged. Results for stations #8 to #11 are shown in figure 4. At stations #1 - #7, no significant differences are observed and the corresponding results are omitted. Computations obtained by using $N_{\text{tot}} = 4$ modes are in good agreement with the measurements only at station #8 and exhibit significant discrepancies at stations #9 to #11. The situation is improved for $N_{\text{tot}} = 5$ modes, which seems sufficient for an adequate reproduction of the experiment. For $N_{\text{tot}} = 7$, the nonlinear dispersive properties of the scheme are further improved leading to excellent agreement with experimental results up to station #11. Results obtained by using $N_{\text{tot}} = 7$ and $N_{\text{tot}} = 10$ modes are practically the same, indicating that convergence has been achieved. The above investigation shows that the dispersive properties of the present formulation are consistently improved by increasing $N_{\text{tot}}$. Similar results are also reported for other models that are based on series representations of polynomial type (Zhao et al., 2014), (Raoult et al., 2016).

### 5.3. *Submerged trapezoid with steep front and back slopes*

We shall now consider more demanding cases where the bathymetry is characterized by larger depth variation and steeper slope. To this aim, the experiment of (Ohyama *et al.*, 1995) is considered, in which incident regular waves are transformed over a submerged isosceles trapezoid of slope 0.5; see Figure 5. Recently, this configuration was used for the validation of the high-order Green-Nagdhi equations (Zhao et al., 2014), the high-order spectral method (Gouin et al., 2016), and a solver for the Navier-Stokes equations with free surface (Chen, Kelly, Dimakopoulos, & Zang, 2016). In the present work, we consider Cases 2, 4 and 6 of

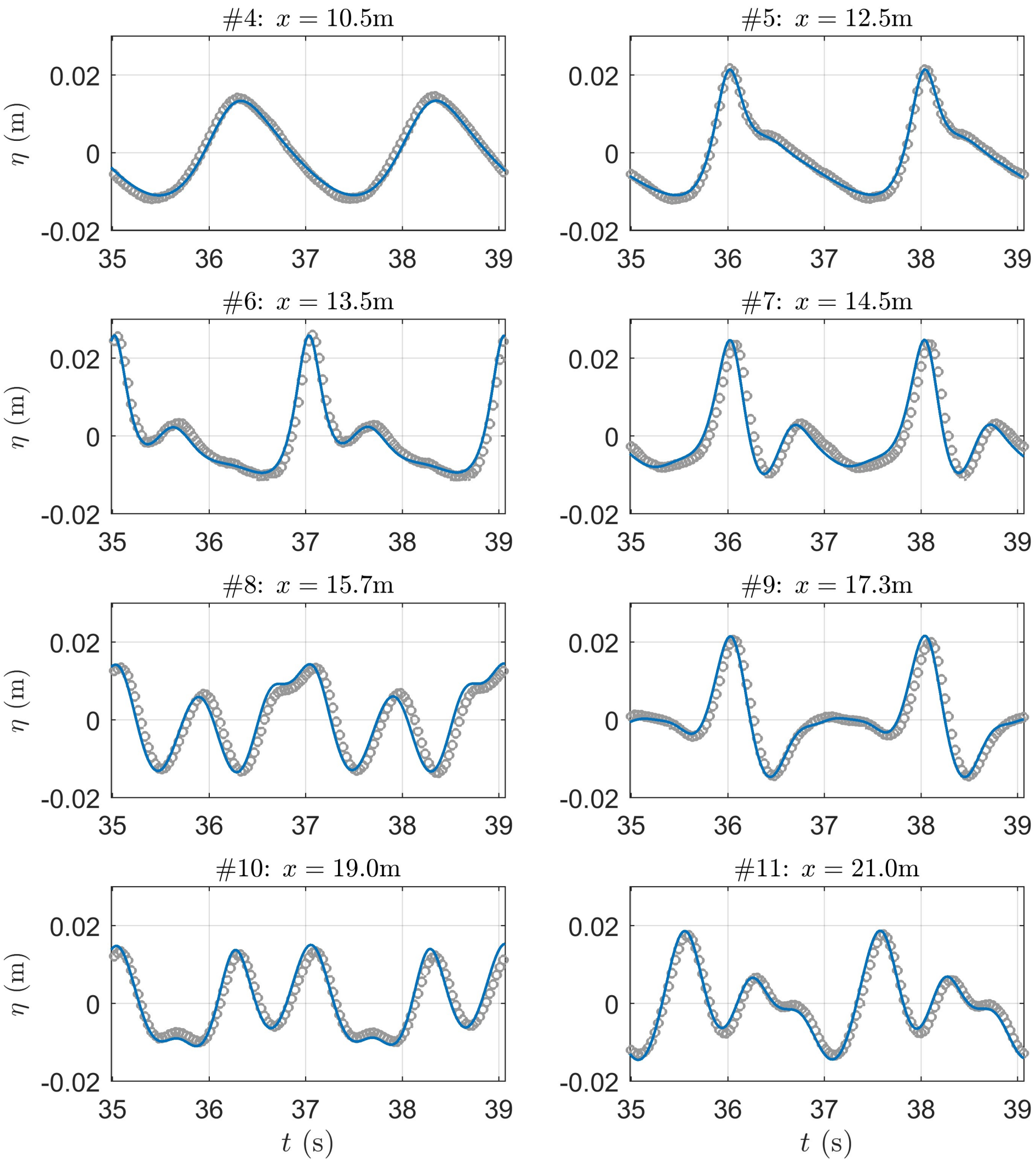

**Figure 3.** Comparison of the computed surface elevation (blue line) with experimental data (circles) of (Dingemans, 1994)

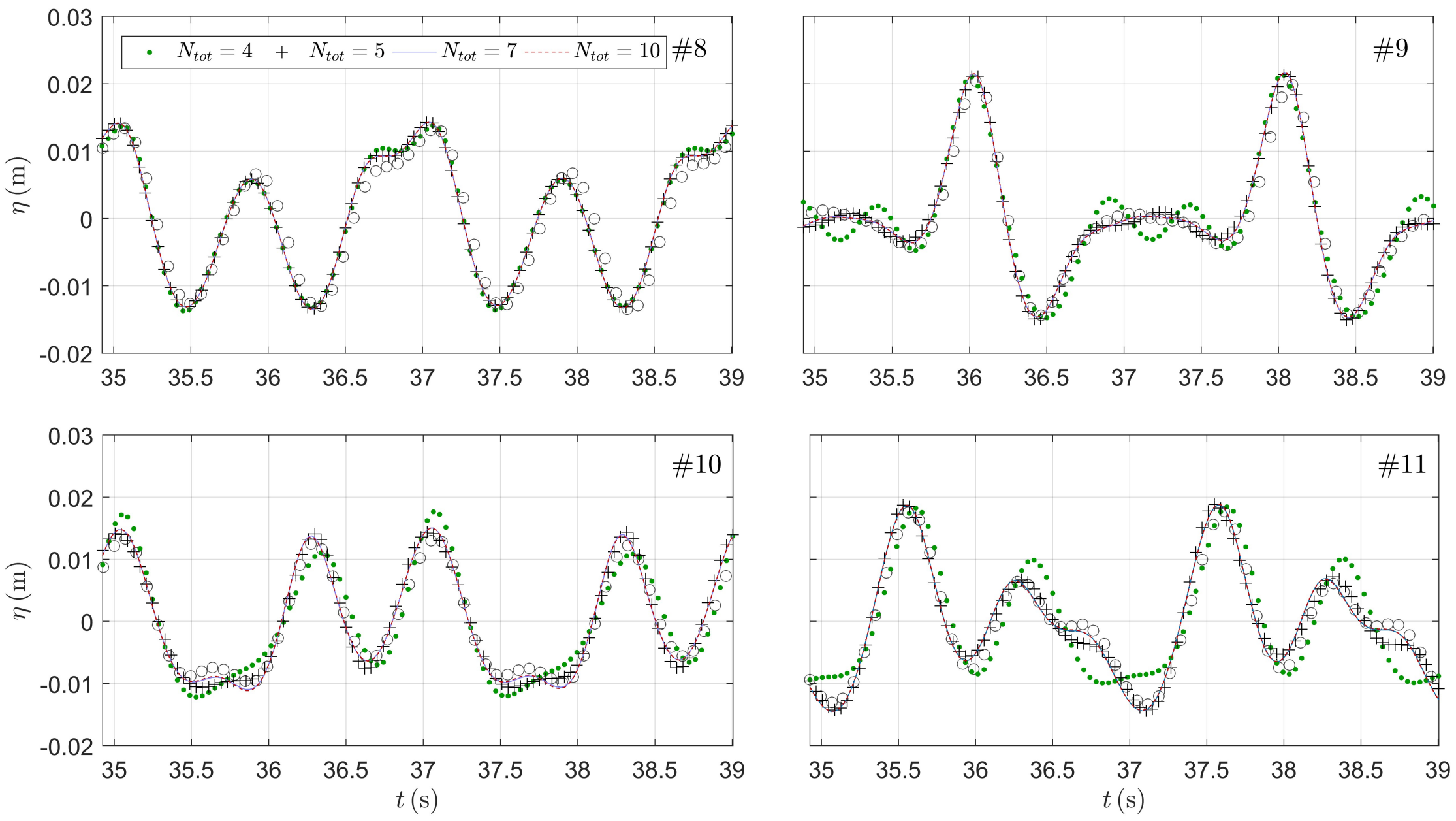

**Figure 4**. Free surface elevation computed by using $N_{\text{tot}} = 4,5,7,10$. Experimental measurements are shown with circles

(Ohyama *et al*., 1995) involving "short", "intermediate" and "long" incident wave conditions, with corresponding heights and periods $(H_0, T_0)$ given by $(0.05, 1.341)$, $(0.05, 2.012)$ and $(0.05, 2.683)$. Denoting by $L_0$ the incident wave length, the spatio-temporal discretization is $\delta x = L_0 / 160 = 0.015$ m, $\delta t = T_0 / 200 = 0.007$ sec for Case 2, $\delta x = L_0 / 200 = 0.02$ m, $\delta t = T_0 / 200 = 0.013$ sec for Case 4, and $\delta x = L_0 / 150 = 0.026$ m, $\delta t = T_0 / 150 = 0.018$ sec for Case 6. Simulations were performed by gradually increasing the total number of modes $N_{\text{tot}}$ until convergence. For all cases, the number $N_{\text{tot}} = 8$ modes was sufficient.

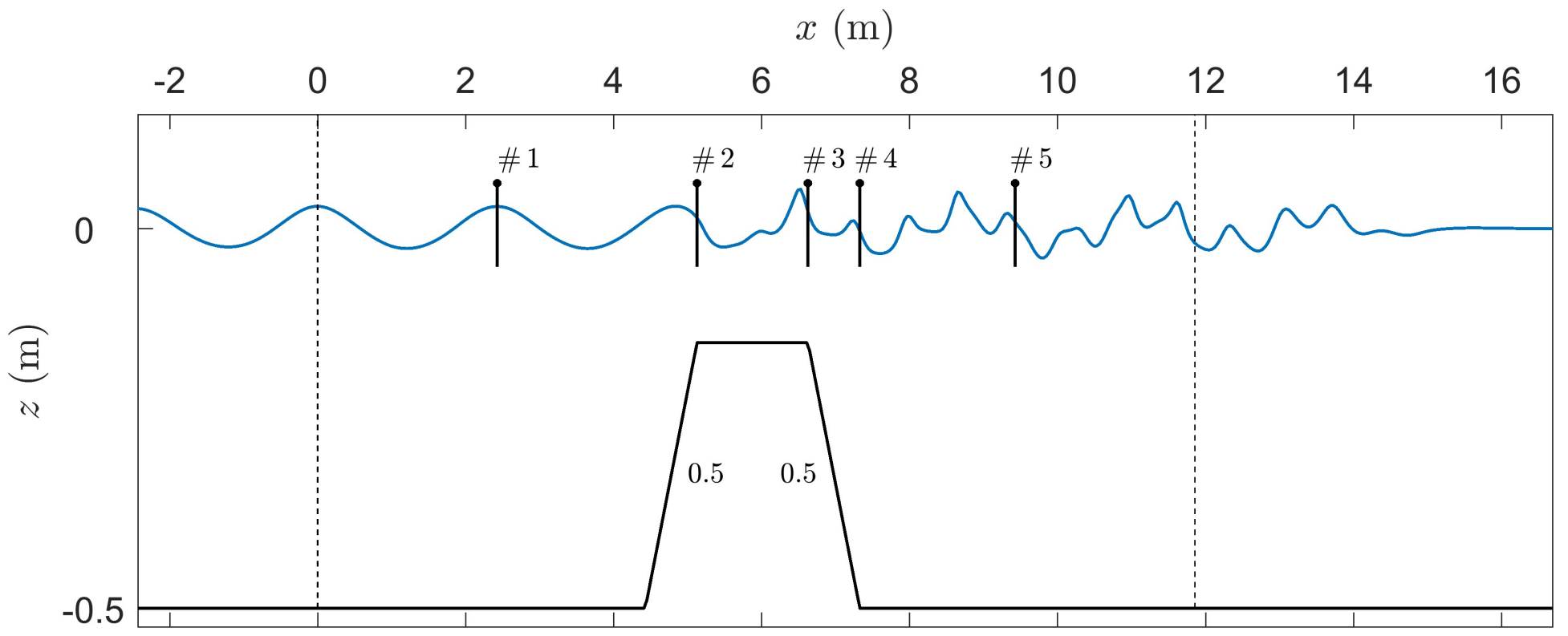

**Figure 5**. Configuration of the numerical wave tank and locations of the measuring stations 1-5 used in (Ohyama *et al*., 1995). Dashed lines correspond to the ends of the sponge layers.

In figure 6, we compare the dimensionless free surface elevation $\eta / H_0$ at stations #3 and #5 with the digitized experimental data and potential flow computations of (Ohyama *et al.*, 1995), obtained by using the Boundary Element Method (BEM). The overall agreement of our computations with the measurements, for all cases at station #3, is very good; in fact, it is better than the agreement of BEM results. A slight difference is observed concerning the maximum free-surface elevation at station #3, which is also observed in the BEM computations of (Ohyama *et al.*, 1995), as well as in the recent results of (Zhao et al., 2014), (Gouin et al., 2016), (Chen et al., 2016). The situation at station #5 is more complicated. Station #5 is located at the lee side of the bar, which is characterized by strongly nonlinear and dispersive effects that cannot be captured by models based on weak nonlinearity and weak dispersion (Ohyama *et al.*, 1995). In contrast, our computations show very good agreement with experimental measurements at Station #5 for Cases 2 and 6. In Case 4, the general trend is adequately captured although a noticeable shift appears in the phase of higher harmonics. This discrepancy is also reported by (Ohyama *et al.*, 1995), (Chen et al., 2016) and (Zhao et al., 2014). (Ohyama *et al.*, 1995) and (Chen et al., 2016) argued that it is likely due to insufficient resolution in their calculations. On the other hand, (Zhao et al., 2014) mention the presence of flow separation or viscosity downstream of the steep bar which cannot be captured by models based on irrotational flow. According to the present authors, a convincing explanation of this discrepancy is lacking.

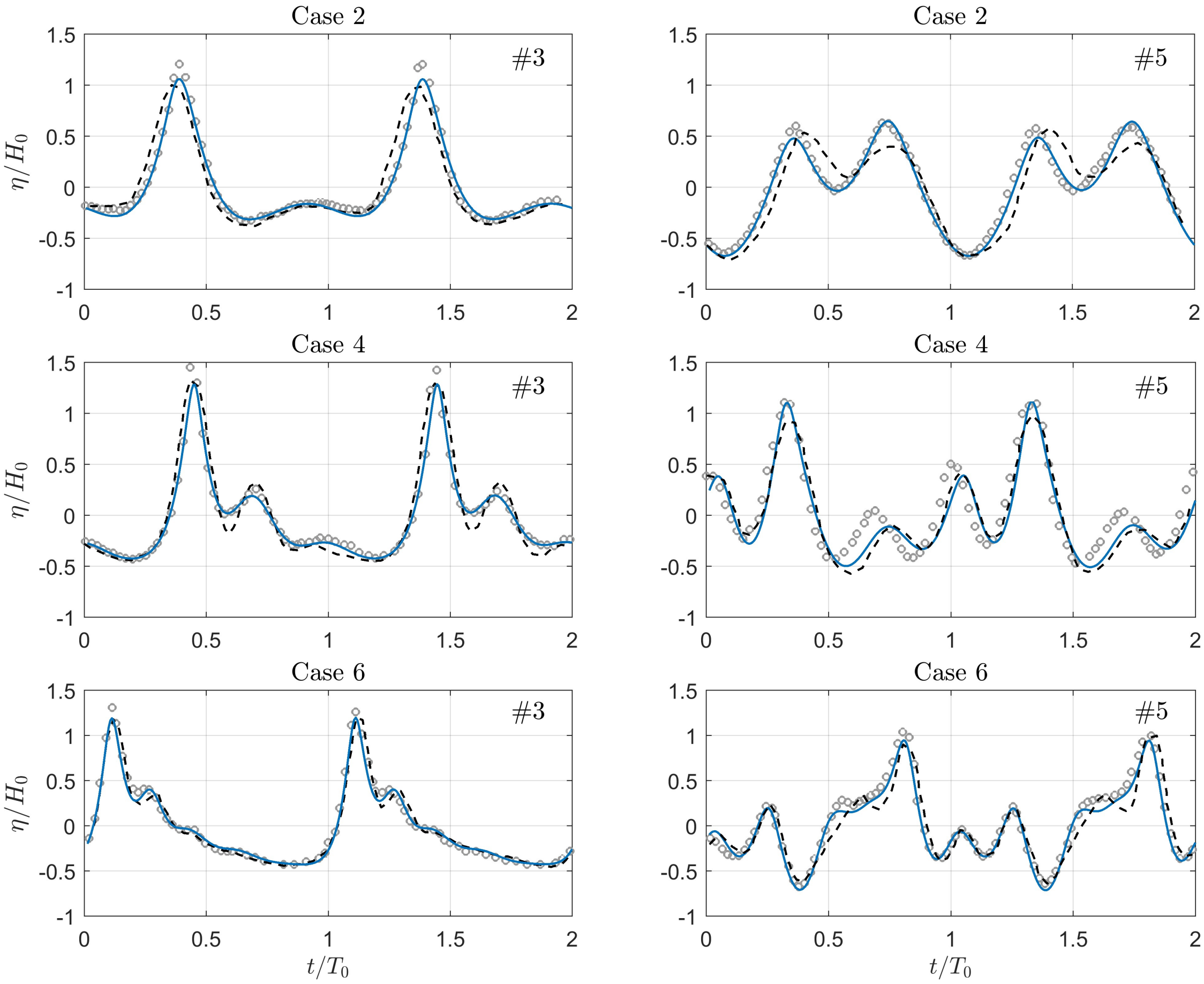

**Figure 6.** Comparison of the computed surface elevation (solid line) with experimental data (circles) and BEM computations (dashed lines) of (Ohyama *et al.*, 1995), for stations #3, #5, and Cases 2,4 and 6.

The results of this section illustrate that the new set of nonlocal, nonlinear evolution equations (4.15) can accurately simulate strongly nonlinear and dispersive waves over strongly varying bathymetry. The accuracy of the proposed scheme is comparable to that of direct numerical methods (e.g. BEM), while it does not require the discretization of the time-dependent free surface or the vertical variable. Further, it avoids the evaluation of high-order polynomial nonlinearities and high-order horizontal spatial derivatives, required in Boussinesq-type models. The overall performance depends on the number of modes/equations kept in the time-independent system (4.12); yet a small number of modes ( $N_{tot} = 6-8$ ) seems to be sufficient even for demanding cases.

## 6. Discussion and conclusions

A new, exact system of nonlocal, nonlinear Hamiltonian evolution equations is derived for the propagation of non-breaking water waves over varying bathymetry, under the assumption of irrotational flow. The derivation is based on Luke's variational principle in conjunction with an exact vertical series expansion of the velocity potential. Apart from smoothness, no further assumptions are made concerning the deformation of the free surface and the bathymetry; that is, the present approach is a *non-pertubative one*. The nonlinear evolution equations are coupled with a system of horizontal partial differential equations that determines the unknown modal amplitudes of the wave potential expansion at every time $t$. Remarkably, only the first modal amplitude, $\varphi_{-2}$, is needed for the computation of the DtN operator that closes the Hamiltonian equations. It is worth noting that the implementation of the DtN operator is equally easy for any (smooth, but arbitrary) bathymetry as for the flat bottom.

Numerical solutions of the new formulation are obtained in the case of one horizontal dimension, by solving the substrate (time-independent) problem using a fourth-order accurate finite-differences scheme, and the Hamiltonian evolution equations by using a classical Runge-Kutta method. In numerical computations the modal series is truncated at a number of $N_{\text{tot}}$ terms. The ignored terms are consistently small, of orders $O(n^{-4})$, independently of how much the boundaries differ from the planar ones. Thus, even the finite-dimensional numerical approximations of the present method *are not perturbative*, as regards the boundary shape. This feature distinguishes the present approach from various other methods which are based on perturbative expansions and, thus, become inefficient and eventually diverge for high values of the slope or amplitude of the boundary deformation. An additional advantage of the present method is that, as shown in AP17, the numerical solver of $\varphi_{-2}$ mode, which represents the DtN operator, exhibits a superconvergence effect, $L^2-\text{Error}[\varphi_{-2}] = O(N_{\text{tot}}^{-6.5})$, making the calculation of the DtN operator very accurate and efficient.

Our computations are validated against measurements from laboratory experiments involving the transformation of incident water waves over submerged obstacles involving mild as well as steep slopes. Excellent agreement with experimental measurements is observed, demonstrating the accuracy and efficiency of the present formulation in simulating coastal flows in the presence of strongly nonlinear and dispersive effects. In all cases only a small number of modes (7 or 8 modes in the most demanding cases) is enough to ensure both numerical convergence of the scheme and nice comparisons with measurements. Thus, the present approach realizes a dimension reduction of the problem, providing results having accuracy and stability comparable with those of direct numerical methods, being much lighter computationally.

The present approach can be easily adapted to cover the case of closed or semi-closed basins with vertical lateral boundaries, as well as the case of waves generated by moving bottom, providing an efficient, fully nonlinear and fully dispersive modeling to tsunamis phenomena. Applications to these directions will be presented elsewhere.

## Appendix A. Proof of Lemma 1

We start by invoking the chain rule for composite operators (see, e.g., (Flett, 1980), 4.1.2, (Gasinski & Papageorgiou, 2005), Prop 4.1.12). Assuming that $G : X \to Y$ has a Gâteaux variation at $x$ for the increment $h$, say $\delta G(x;h)$, and $F : Y \to Z$ is Fréchet differentiable at $y = G(x)$, with Fréchet derivative $DF(y)$, the composition $W = F \circ G$ has a Gâteaux variation at $x$ for the increment $h$, $\delta W(x;h)$, given by

$$\delta W(x;h) = DF(y)\big|_{y=G(x)} \left[\delta G(x;h)\right] = \delta F(G(x);\delta G(x;h)). \qquad \text{(A1)}$$

In our case $x = (x_1, x_2) = (\eta, \boldsymbol{\varphi})$, $y = (y_1, y_2) = (\eta, \Phi)$, $W$ and $F$ are (the action) functionals on $(x_1, x_2)$ and $(y_1, y_2)$, respectively, and $G$ is the operator transforming $(x_1, x_2)$ to $(y_1, y_2)$, having the form $(y_1, y_2) = G(x_1, x_2) = \left(G_1(x_1, x_2), G_2(x_1, x_2)\right)$. Then, functional $W$ takes the form $W(x_1, x_2) = F(G(x_1, x_2)) = F(G_1(x_1, x_2), G_2(x_1, x_2))$. The partial variations $\delta_{x_1} W(x;h_1)$ and $\delta_{x_2} W(x;h_2)$ are calculated by applying the chain rule (A1) separately, in the directions $h_1$ and $h_2$:

$$\delta_{x_1} W(x;h_1) = \delta_{y_1} F(G(x);\delta_{x_1} G_1(x;h_1)) + \delta_{y_2} F(G(x);\delta_{x_1} G_2(x;h_1)), \qquad \text{(A2a)}$$

$$\delta_{x_2} W(x;h_2) = \delta_{y_1} F(G(x);\delta_{x_2} G_1(x;h_2)) + \delta_{y_2} F(G(x);\delta_{x_2} G_2(x;h_2)). \qquad \text{(A2b)}$$

For the specific application considered herewith, the composite functional is $W(x_1, x_2) \equiv \tilde{\mathcal{S}}[\eta, \boldsymbol{\varphi}] = \mathcal{S}[G_1(\eta, \boldsymbol{\varphi}), G_2(\eta, \boldsymbol{\varphi})]$, the initial functional $F(y_1, y_2)$ is $\mathcal{S}[\eta, \Phi]$, and $G$ is the operator transforming $[\eta, \boldsymbol{\varphi}]$ to $[\eta, \Phi]$, with $G_1(\eta, \boldsymbol{\varphi}) = \eta$ and $G_2(\eta, \boldsymbol{\varphi}) = \Phi(\eta, \boldsymbol{\varphi})$, the latter being expressed by means of the series representation (3.1), i.e., $\Phi(\eta, \boldsymbol{\varphi}) = \boldsymbol{\varphi}^{\mathrm{T}} \boldsymbol{Z}(\eta)$. In order to apply the results (A2a,b) in our case, the partial variations of $G_1(\eta, \boldsymbol{\varphi})$ and $G_2(\eta, \boldsymbol{\varphi})$ are needed. The calculation of the partial variations of $G_1(\eta, \boldsymbol{\varphi}) = \eta$ is trivial: $\delta_\eta G_1(\eta, \boldsymbol{\varphi}; \delta\eta) = \delta\eta$, $\delta_{\boldsymbol{\varphi}} G_1(\eta, \boldsymbol{\varphi}; \delta\boldsymbol{\varphi}) = \boldsymbol{0}$ (since $\eta$ and $\boldsymbol{\varphi}$ are independent functional arguments). Concerning $G_2(\eta, \boldsymbol{\varphi})$, we observe that it is linearly dependent on $\boldsymbol{\varphi} = \{\varphi_m\}_m$, and nonlinearly but explicitly dependent on $\eta$, through the prescribed functions $\boldsymbol{Z}(\eta)$. It follows that its partial variations are given by

$$\delta_\eta G_2(\eta, \boldsymbol{\varphi}; \delta\eta) = \delta_\eta \Phi(\eta, \boldsymbol{\varphi}; \delta\eta) = \partial_\eta \Phi(\eta, \boldsymbol{\varphi})\, \delta\eta, \qquad \text{(A3a)}$$

and

$$\delta_{\boldsymbol{\varphi}} G_2(\eta, \boldsymbol{\varphi}; \delta\boldsymbol{\varphi}) = \delta_{\boldsymbol{\varphi}} \Phi(\eta, \boldsymbol{\varphi}; \delta\boldsymbol{\varphi}) = \partial_{\boldsymbol{\varphi}} \Phi(\eta, \boldsymbol{\varphi})^{\mathrm{T}}\, \delta\boldsymbol{\varphi}. \qquad \text{(A3b)}$$

Note that the last members of the above equations are expressed in terms of usual partial derivatives of $\Phi(\eta,\boldsymbol{\varphi})$, given by $\partial_\eta \Phi = \boldsymbol{\varphi}^{\mathrm{T}} \partial_\eta \boldsymbol{Z}(\eta)$ and

$$\partial_{\boldsymbol{\varphi}} \Phi(\eta,\boldsymbol{\varphi}) = \left(\partial_{\varphi_1}\Phi, \partial_{\varphi_2}\Phi, \dots, \partial_{\varphi_m}\Phi, \dots\right) = \left(Z_1, Z_2, \dots, Z_m, \dots\right) = \boldsymbol{Z}.$$

The partial variations $\delta_\eta \tilde{\mathcal{S}}[\eta,\boldsymbol{\varphi};\delta\eta]$ and $\delta_{\boldsymbol{\varphi}} \tilde{\mathcal{S}}[\eta,\boldsymbol{\varphi};\delta\boldsymbol{\varphi}]$ are obtained by applying the chain rule, Eqs. (A2a,b), and using Eqs. (A3a,b),

$$\delta_\eta \tilde{\mathcal{S}}[\eta,\boldsymbol{\varphi};\delta\eta] = \delta_\eta \mathcal{S}[\eta,\boldsymbol{\varphi}^{\mathrm{T}}\boldsymbol{Z}(\eta);\delta\eta] + \delta_\Phi \mathcal{S}[\eta,\boldsymbol{\varphi}^{\mathrm{T}}\boldsymbol{Z}(\eta);(\boldsymbol{\varphi}^{\mathrm{T}}\partial_\eta \boldsymbol{Z}(\eta))\delta\eta], \quad \text{(A4a)}$$

$$\delta_{\boldsymbol{\varphi}} \tilde{\mathcal{S}}[\eta,\boldsymbol{\varphi};\delta\boldsymbol{\varphi}] = \delta_\Phi \mathcal{S}[\eta,\boldsymbol{\varphi}^{\mathrm{T}}\boldsymbol{Z}(\eta);\boldsymbol{Z}(\eta)^{\mathrm{T}}\delta\boldsymbol{\varphi}]. \quad \text{(A4b)}$$

Eq. (A4a) is identical to Eq. (3.4a). Invoking the linearity of the variation with respect to the increment, i.e. $\delta_{\boldsymbol{\varphi}} \tilde{\mathcal{S}}[\eta,\boldsymbol{\varphi};\delta\boldsymbol{\varphi}] = \sum_m \delta_{\varphi_m} \tilde{\mathcal{S}}[\eta,\boldsymbol{\varphi};\delta\varphi_m]$, and observing that Fréchet differentiability of $\mathcal{S}[\eta,\Phi]$ implies that $\delta_\Phi \mathcal{S}[\eta,\Phi;\bullet]$ is a linear operator on $\delta\Phi$, we can write $\delta_{\boldsymbol{\varphi}} \tilde{\mathcal{S}}[\eta,\boldsymbol{\varphi};\boldsymbol{\delta\varphi}]$ and $\boldsymbol{Z}(\eta)^{\mathrm{T}}\delta\boldsymbol{\varphi}$ as sums in Eq. (A4b), to obtain $\sum_m \delta_{\varphi_m} \tilde{\mathcal{S}}[\eta,\boldsymbol{\varphi};\delta\varphi_m] = \sum_m \delta_\Phi \mathcal{S}[\eta,\boldsymbol{\varphi}^{\mathrm{T}}\boldsymbol{Z}(\eta);Z_m(\eta)\delta\varphi_m]$; thus, Eq. (3.4b) follows by using the arbitrariness of the increments $\delta\varphi_m$ and the proof is complete.